\documentclass[aps,prc,twocolumn,groupedaddress]{revtex4-1}
\usepackage{graphicx}
\usepackage{amsmath}

\begin{document}

% Use the \preprint command to place your local institutional report
% number in the upper righthand corner of the title page in preprint mode.
% Multiple \preprint commands are allowed.
% Use the 'preprintnumbers' class option to override journal defaults
% to display numbers if necessary
%\preprint{}

%Title of paper
\title{Total Absorption $\gamma$-Ray Spectroscopy of $^{87}$Br, $^{88}$Br and $^{94}$Rb $\beta$-Delayed Neutron Emitters}

% repeat the \author .. \affiliation  etc. as needed
% \email, \thanks, \homepage, \altaffiliation all apply to the current
% author. Explanatory text should go in the []'s, actual e-mail
% address or url should go in the {}'s for \email and \homepage.
% Please use the appropriate macro foreach each type of information

% \affiliation command applies to all authors since the last
% \affiliation command. The \affiliation command should follow the
% other information
% \affiliation can be followed by \email, \homepage, \thanks as well.
%\author{}
%\homepage[]{Your web page}
%\thanks{}
%\altaffiliation{}
\author{E. Valencia}
\author{J. L. Tain}
\email[Corresponding author:]{tain@ific.uv.es}
\author{A. Algora}
\altaffiliation{Institute of Nuclear Research of the Hungarian Academy of Sciences, H-4026 Debrecen, Hungary}
\author{J. Agramunt}
\author{E. Estevez}
\author{M.D. Jordan}
\author{B. Rubio}
\affiliation{Instituto de Fisica Corpuscular (CSIC-Universitat de Valencia),
  Apdo. Correos 22085, E-46071 Valencia, Spain}

\author{S. Rice}
\author{P. Regan}
\author{W. Gelletly}
\author{Z. Podoly\'ak}
\author{M. Bowry}
\author{P. Mason}
\author{G. F. Farrelly}
\affiliation{University of Surrey, Department of Physics, Guilford GU2 7XH, United Kingdom}

\author{A. Zakari-Issoufou}
\author{M. Fallot}
\author{A. Porta}
\author{V. M. Bui}
\affiliation{SUBATECH, CNRS/IN2P3, Université de Nantes, Ecole des Mines, F-44307 Nantes, France}

\author{J. Rissanen}
\author{T. Eronen}
\author{I. Moore}
\author{H. Penttil\"a}
\author{J. \"Ayst\"o}
\author{V.-V. Elomaa}
\author{J. Hakala}
\author{A. Jokinen}
\author{V. S. Kolhinen}
\author{M. Reponen}
\author{V. Sonnenschein}
\affiliation{University of Jyvaskyla, Department of Physics, P.O. Box 35, FI-40014 University of Jyvaskyla, Finland}

\author{D. Cano-Ott}
\author{A. R. Garcia}
\author{T. Mart\'{\i}nez}
\author{E. Mendoza}
\affiliation{Centro de Investigaciones Energ\'eticas Medioambientales y Tecn\'ologicas, E-28040 Madrid, Spain}

\author{R. Caballero-Folch}
\author{B. Gomez-Hornillos}
\author{V. Gorlichev}
\affiliation{Universitat Politecnica de Catalunya, E-08028 Barcelona, Spain}

\author{F. G. Kondev}
\affiliation{Nuclear Engineering Division, Argonne National Laboratory, Argonne, Illinois 60439, USA}

\author{A. A. Sonzogni}
\affiliation{NNDC, Brookhaven National Laboratory, Upton, New York 11973, USA}

\author{L. Batist}
\affiliation{Petersburg Nuclear Physics Institute, Gatchina, Russia}

%Collaboration name if desired (requires use of superscriptaddress
%option in \documentclass). \noaffiliation is required (may also be
%used with the \author command).
%\collaboration can be followed by \email, \homepage, \thanks as well.
%\collaboration{}
%\noaffiliation

\date{\today}

\begin{abstract}
% insert abstract here
We investigate the decay of $^{87,88}$Br and $^{94}$Rb using total absorption $\gamma$-ray spectroscopy.
These important fission products are $\beta$-delayed neutron emitters.
Our data show considerable $\beta\gamma$-intensity, so far unobserved
in high-resolution $\gamma$-ray spectroscopy, from states at high excitation energy.
We also find significant differences with the $\beta$ intensity that can be deduced
from existing measurements of the $\beta$ spectrum.
We evaluate the impact of the present data on reactor decay heat using
summation calculations. Although the effect is relatively small it helps to reduce the discrepancy 
between calculations and integral measurements of the photon component for $^{235}$U fission
at cooling times in the range $1-100$~s.
We also use summation calculations to evaluate the impact of present data on 
reactor antineutrino spectra. We find a significant effect at antineutrino energies
in the range of 5 to 9~MeV. In addition, we observe an unexpected 
strong probability  for $\gamma$ emission
from neutron unbound states populated in the daughter nucleus. The $\gamma$ branching
is compared to Hauser-Feshbach calculations which allow one to explain the  
large value for bromine isotopes as due to nuclear structure. However the branching
for $^{94}$Rb, although much smaller, hints of the need to increase the radiative 
width $\Gamma_{\gamma}$ by one order-of-magnitude. This leads to a similar increase
in the calculated $(\mathrm{n},\gamma)$ cross section for this
very neutron-rich nucleus with a potential impact on $r$ process abundance calculations. 
\end{abstract}

% insert suggested PACS numbers in braces on next line
%\pacs{}
% insert suggested keywords - APS authors don't need to do this
%\keywords{}

%\maketitle must follow title, authors, abstract, \pacs, and \keywords
\maketitle

\section{Introduction
\label{intro}}

Neutron-unbound states can be populated in the $\beta$-decay of
very neutron-rich nuclei, when the neutron separation energy
$S_{n}$ in the daughter nucleus is lower than the decay energy window $Q_{\beta}$.
The relative strength of strong and electromagnetic interactions determines the preponderance
of neutron emission over
$\gamma$-ray emission from these states. These emission rates are
quantified by the partial level widths $\Gamma_{n}$ and $\Gamma_{\gamma}$
respectively. The fraction of $\beta$ intensity followed by $\gamma$-ray
emission is given by 
$\Gamma_{\gamma} / \Gamma_{tot}$, with $\Gamma_{tot} =\Gamma_{\gamma} + \Gamma_{n}$.
There is an analogy~\cite{kra83} between this decay process and 
neutron capture reactions populating unbound
states. Such resonances in the compound nucleus re-emit a neutron (elastic channel) or
de-excite by $\gamma$-rays (radiative capture). Indeed the reaction
cross section is parametrized in terms of neutron and $\gamma$ widths.
In particular the $(\mathrm{n},\gamma)$ cross section includes terms
proportional to  $\Gamma_{\gamma} \Gamma_{n} / \Gamma_{tot}$.
Notice that the spins and parities of states populated
in $\beta$-decay and  $(\mathrm{n},\gamma)$ do not coincide in general because of the different
selection rules.

Neutron capture and transmission reactions have been extensively used~\cite{mug06}
to determine $\Gamma_{\gamma}$ and $\Gamma_{n}$ of resolved resonances,
or the related strength functions in
the unresolved resonance region.
An inspection of Ref.~\cite{mug06} shows
that in general $\Gamma_{n}$ is measured in eV or keV
while $\Gamma_{\gamma}$ is measured in meV or eV, in agreement with expectation.
Current data is restricted, however, to nuclei close
to stability since such experiments require the use of stable or long-lived targets. 
On the other hand, $(\mathrm{n},\gamma)$ capture 
cross sections for very neutron-rich nuclei are 
a key ingredient in reaction network calculations describing the synthesis 
of elements heavier than iron during the rapid ($r$) neutron capture
process occurring in explosive-like stellar events. 
In the classical picture of the $r$ process~\cite{b2fh}
a large burst of neutrons synthesizes the elements along a path determined
by the $(\mathrm{n},\gamma)-(\gamma,\mathrm{n})$ equilibrium. 
After the exhaustion of neutrons 
these isotopes decay back to the $\beta$-stability valley. In this simplified
model the capture cross section magnitude plays no role. 
However it is known~\cite{gor98,sur01,arc11} that for
realistic irradiation scenarios the final elemental abundance is sensitive to the actual
$(\mathrm{n},\gamma)$ cross-sections. This is the case for the hot (classical) $r$ process,
due to the role of late captures during the decay back to stability.
It is also the case for a cold r-process, where the formation path is determined 
by competition between neutron capture and beta decay.

Lacking experimental information, the cross section for these exotic nuclei
is typically obtained from Hauser-Feshbach statistical model calculations~\cite{rau00}. 
This model is based on a few quantities describing average properties of the nucleus: 
the nuclear level density (NLD), the photon strength
function (PSF) and the neutron transmission coefficient (NTC). 
The PSF determines $\Gamma_{\gamma}$, NTC determines $\Gamma_{n}$ and
NLD affects both (see Appendix).
The parameters describing the dependence of these quantities on various magnitudes
are adjusted to experiment close to $\beta$ stability.
It is thus crucial to find means to verify the predictions of the model far from
stability. 
For example, the use of surrogate reactions with radioactive beams has been 
suggested as a tool to provide experimental constraints on 
$(\mathrm{n},\gamma)$ cross sections~\cite{esc12} for unstable nuclei, 
but its application is very challenging and restricted to nuclei close to stability.
On the other hand the study of $\gamma$-ray emission from states above $S_{n}$ 
observed in $\beta$ decay can give quantitative information on 
$\Gamma_{\gamma} / \Gamma_{tot}$ for unstable nuclei. This
information can be used to improve neutron capture cross-section estimates
for nuclei far away from $\beta$ stability.

The emission of $\gamma$ rays from neutron unbound states populated in
$\beta$ decay has been observed in very few cases studied with high-resolution 
germanium detectors. 
It was first detected in 1972 in the decay of $^{87}$Br~\cite{sla72}
which remains one of the best studied cases~\cite{tov75,nuh77,ram83}. 
The other cases are:
$^{137}$I~\cite{nuh75,kra79,ohm80}, $^{93}$Rb~\cite{bis77,alk89,gre97}, 
$^{85}$As~\cite{kra79,omt91}, 
$^{141}$Cs~\cite{yam82}, $^{95}$Rb~\cite{kra83a}, $^{94}$Rb~\cite{alk89},
$^{77}$Cu~\cite{ily09}, and $^{75}$Cu~\cite{ily11}. 
In the decay of $^{87}$Br up to a dozen states 
emitting single $\gamma$-rays
have been identified within 250~keV above $S_{n}$,
with a total intensity of  about 0.5\% compared with a neutron emission intensity of 2.6\%. 
The observation of relatively intense $\gamma$-rays in this measurement was
explained as being due to nuclear structure since some of the
levels populated could only decay through the hindered emission of a high orbital angular 
momentum neutron. 
On the other hand, it was pointed out~\cite{jon76} that a sizable $\gamma$-ray
emission from neutron unbound states could be a manifestation of Porter-Thomas (PT)
statistical fluctuations in the strength of individual transitions.
The extremely asymmetric shape of the PT distribution can lead to
very large enhancement of the $\Gamma_{\gamma} / \Gamma_{tot}$ ratio
with respect to the average.
However a general characterization of the phenomenon is still lacking,
in particular the relative importance of the different mechanisms governing the competition.

It is difficult to pursue these studies using high-resolution $\gamma$-ray spectroscopy.
Since the $\beta$ intensity is distributed over many highly excited states, 
germanium detectors are prone to miss the $\gamma$-ray de-excitation intensity, further
fragmented over multiple cascades. This has come to be known
as the {\it Pandemonium} effect~\cite{har77}. 
Therefore in order to measure the full $\gamma$ intensity one 
must resort to the Total Absorption Gamma-ray Spectroscopy (TAGS) technique  
aimed at detecting cascades rather than individual $\gamma$ rays. 
The power of this method to locate missing
$\beta$ intensity has been
demonstrated before~\cite{alg99,alg03,hu99}. However its application in the present case is 
very challenging, since the expected $\gamma$ branching is very small. 
As a matter of fact previous attempts at the Leningrad Nuclear Physics Institute (LNPI)~\cite{alk89}  
did not lead to clear conclusions. 
In this work we propose and demonstrate for the first time the use of the TAGS technique
to study $\beta$-delayed neutron emitters and extract accurate information
that can be used to constrain $(\mathrm{n},\gamma)$ 
cross-section estimates for very unstable nuclei.

For this study we selected three known
neutron emitters: $^{87}$Br, $^{88}$Br and $^{94}$Rb.
The relevant decay parameters are given in Table~\ref{isotdata}.
The quoted quantities: $T_{1/2}$ (half life), $P_{n}$ (neutron emission probability), 
$Q_{\beta}$ and $S_{n}$ are taken from the ENSDF data base~\cite{br87,br88,rb94}.
The case of $^{87}$Br has been included in the study on purpose since
it allows a comparison of our results with high resolution decay 
measurements~\cite{ram83} and with neutron capture and 
transmission experiments~\cite{ram83,car88}.
The case of $^{93}$Rb was also measured and will be presented separately~\cite{zak15a}.

%%%%%  Table 1  %%%%%
\begin{table}[h]
\caption{Half-life $T_{1/2}$, neutron emission probability $P_{n}$, 
decay energy window $Q_{\beta}$, and daughter neutron separation energy $S_{n}$
for each measured isotope. Values taken from Ref.~\cite{br87,br88,rb94}.}
\begin{center}
%\resizebox{5.7cm}{!}{
\begin{tabular}{ccccc} \hline \hline
 & $T_{1/2}$ & $P_{n}$ & $Q_{\beta}$ & $S_{n}$ \\
Isotope & (s) & (\%) & (MeV) & (MeV) \\ \hline
$^{87}$Br & 55.65(13) & 2.60(4) & 6.852(18) & 5.515(1) \\
$^{88}$Br & 16.34(8) &  6.58(18) & 8.975(4) & 7.054(3) \\
$^{94}$Rb & 2.702(5) & 10.18(24) & 10.281(8) & 6.828(10) \\ \hline \hline
\end{tabular}
%}
\end{center}
\label{isotdata}
\end{table}

The results of our TAGS analysis are also relevant for 
reactor decay heat and anti-neutrino spectrum
calculations.

The knowledge of the heating produced by radioactive products in a reactor 
and the time evolution
after reactor shutdown is important for reactor safety.
In conventional reactors the decay heat (DH) is dominated by fission products (FP)
for cooling times up to a few years.
An issue in reactor DH studies
has been the persistent failure of summation calculations
to reproduce the results of integral experiments
for individual fissioning systems. 
Summation calculations are based on individual FP yields and
average $\gamma$-ray and $\beta$ energies retrieved from evaluated nuclear data bases.
In spite of this deficiency
summation calculations remain an important tool in reactor safety studies.
For example, after the Fukushima Dai-ichi nuclear plant accident 
it was pointed out~\cite{oku13} that summation calculations 
are relevant to understand  the progression
of core meltdown in this type of event.
The Fukushima accident was the consequence
of a failure to dissipate effectively the DH in the reactor core and 
in the adjacent spent fuel cooling pool. 
Summation calculations are particularly important in design studies of innovative reactor
systems (Gen IV reactors, Accelerator Driven Systems) with unusual fuel compositions 
(large fraction of minor actinides), high burn ups and/or harder neutron spectra,
since integral data are missing.

Yoshida and Nakasima~\cite{yos81} recognized that the $Pandemonium$
systematic error  is responsible
for a substantial fraction of the discrepancy between DH integral experiments and calculations. 
\emph{Pandemomium} has the effect of decreasing the average $\gamma$-ray  energy
and increasing the average $\beta$ energy when calculated
from available level scheme information obtained in high resolution measurements.
The average $\gamma$ and $\beta$ energy for each isotope, 
$\bar{E}_{\gamma}$ and $\bar{E}_{\beta} $ respectively,
can be computed from $I_{\beta} (E_{x})$, 
the $\beta$ intensity distribution as a function of excitation energy $E_{x}$ as

\begin{eqnarray}
%\begin{gathered}
\bar{E}_{\gamma} = \int_{0}^{Q_{\beta}} I_{\beta} (E_{x}) E_{x} dE_{x} 
\label{eq:avereneg} \\
\bar{E}_{\beta} = \int_{0}^{Q_{\beta}} I_{\beta} (E_{x}) \langle E_{\beta} (Q_{\beta}-E_{x}) \rangle dE_{x} 
\label{eq:avereneb}
%\end{gathered}
\end{eqnarray}

Here $\langle E_{\beta} (Q_{\beta}-E_{x}) \rangle$ represents the mean value of the $\beta$ energy continuum
leading to a state at $E_{x}$.

The TAGS technique, free from $Pandemonium$,
was applied in the 1990s by Greenwood and collaborators at INEL (Idaho)~\cite{gre97}
to obtain accurate average decay energies for up to 48 FP
with impact in DH calculations.
Recognizing the importance of this approach to improve summation calculations,
the OECD/NEA Working Party on International Evaluation Cooperation (WPEC)
established subgroup SG25 to review the situation~\cite{sg25}. 
They made recommendations, in the form of priority lists,
for future TAGS measurements on specific isotopes for the U/Pu fuel cycle. 
The work was later extended to the Th/U fuel cycle by Nichols and collaborators~\cite{gup10}.
Subsequently, the results of Algora {\it et al.}~\cite{alg10} demonstrated the large impact 
of new TAGS measurements for a few isotopes selected 
from the priority list.
 
From the nuclei included in the present work
$^{87}$Br was assigned priority 1 in Ref.~\cite{sg25,gup10}
for a TAGS measurement, although 
it is an example of a well studied
level scheme~\cite{br87} with  up to 374 $\gamma$ transitions de-exciting 
181 levels. The justification for the high priority comes from:  
1) the large uncertainty (25\%) on average energies coming from the spread of 
intensity normalization values between different measurements,  2) a potential $Pandemonium$
error suggested by the number of observed levels at high excitation energies
(less than half of the expected number according to level density estimates), 
and 3) the large contribution
to DH around 100~s cooling time. $^{88}$Br also has priority 1
in Ref.~\cite{sg25,gup10}. It contributes significantly to the DH at cooling times
around 10~s. The known decay scheme~\cite{br88} is rather incomplete
above $E_{x}=3.5$~MeV,  from level density considerations, as
shown in the RIPL-3 reference input parameter library web page~\cite{ripl3}.
We estimate that more than 300 
levels should be populated in the decay above $E_{x}=3.5$~MeV and below $S_{n}$
in comparison with the observed number of 33.
$^{94}$Rb is not included in the priority list of Ref.~\cite{sg25} 
but is considered of relative importance in Refs.~\cite{gup10} and \cite{fle15} for
short cooling times. The decay scheme is very poorly known~\cite{rb94}.
Only 37 levels are identified above $E_{x}=3.4$~MeV, regarded as the maximum energy with
a complete level scheme~\cite{ripl3}. 
We estimate that more than 900 levels could be populated
below $S_{n}$ thus pointing to a potentially strong $Pandemonium$ effect.

Summation calculations are also a valuable tool to study reactor anti-neutrino 
$\bar{\nu}_{e}$ spectra.
Accurate knowledge of this spectrum is of relevance for the analysis of neutrino
oscillation experiments~\cite{bem02,kim13} and for exploring the use of compact anti-neutrino
detectors in nuclear proliferation control~\cite{cri11}.
Summation calculations for the $\bar{\nu}_{e}$ spectrum suffer from the same problem
as DH summation calculations:
inaccuracies in fission yields and individual precursor decay data.

For each fission product  the electron antineutrino spectrum $S_{\bar{\nu}} (E_{\bar{\nu}})$, 
and the related $\beta$ spectrum $S_{\beta} (E_{\beta})$, 
can be computed  from the $\beta$ intensity distribution

\begin{eqnarray}
S_{\bar{\nu}} (E_{\bar{\nu}}) = \int_{0}^{Q_{\beta}} I_{\beta} (E_{x}) s_{\bar{\nu}} (Q_{\beta}-E_{x}, E_{\bar{\nu}})  dE_{x} \label{eq:nuspec} \\
S_{\beta} (E_{\beta}) = \int_{0}^{Q_{\beta}} I_{\beta} (E_{x}) s_{\beta} (Q_{\beta}-E_{x}, E_{\beta})  dE_{x} 
\label{eq:betspec}
\end{eqnarray}

where $s_{\bar{\nu}} (Q_{\beta}-E_{x}, E_{\bar{\nu}})$ and $s_{\beta} (Q_{\beta}-E_{x}, E_{\beta})$ 
represent the shape of $\bar{\nu}_{e}$ and $\beta$ energy distributions for
the transition to a state at $E_{x}$, which depends on the nuclear wave functions.
For each $E_{x}$, $s_{\bar{\nu}}$ and $s_{\beta}$ are related by energy-conservation
$E_{\bar{\nu}} = Q_{\beta}-E_{x}-E_{\beta}$ to a good approximation~\cite{mou15}.
Thus distortions of the observed $I_{\beta} (E_{x})$ distribution 
in high-resolution $\gamma$-ray spectroscopy due to $Pandemonium$
tend to produce calculated $\bar{\nu}_{e}$ spectra shifted to high energies.

Currently the most reliable reactor $\bar{\nu}_{e}$ spectra are obtained from integral
$\beta$-spectrum measurements of $^{235}$U, $^{239}$Pu and $^{241}$Pu thermal fission 
performed by Schreckenbach {\it et al.} at ILL-Grenoble~\cite{sch85,hah89}.  Data 
on $^{238}$U fast fission also became available recently~\cite{haa14}. 
The conversion of integral $\beta$ spectra to $\bar{\nu}_{e}$ spectra requires a number of
approximations. These are needed because, as pointed out above, 
the transformation is isotope and level dependent. 
The global conversion procedure has been revised and improved recently~\cite{mue11,hub11}.
As a consequence of this revision a change of normalization in the spectrum is found
that contributes to a consistent deficit when comparing $\bar{\nu}_{e}$ rates from 
short base line experiments with
calculations \cite{men11}, a surprising effect which is termed the reactor neutrino anomaly.
The effect has been related to an abundance of transitions of the first forbidden type
with a $\beta$ spectrum departing from the allowed shape~\cite{fan15}.
The experimental investigation of this or similar effects requires accurate measurements of
individual fission products and the use of the summation method
as was argued in \cite{fal12}.

The statistics accumulated in the three running reactor $\bar{\nu}_{e}$ experiments,
Double Chooz~\cite{abe14}, RENO~\cite{cho16} and Daya Bay~\cite{an16},
has revealed differences between the  shape 
of the calculated $\bar{\nu}_{e}$ spectra and the measured one.
The observed excess between 5 and 7~MeV $E_{\bar{\nu}_{e}}$ could be due to the 
contribution of a few
specific FP~\cite{dwy15,son15} which is not reproduced by the global
conversion method. 
Thus the study of this new antineutrino shape distortion 
requires the use of the summation method and reinforces the need for new accurate
decay data with the TAGS technique. As a matter of fact 
one of the key isotopes in this list, $^{92}$Rb, was part of the same experiment
analyzed here and its impact on the antineutrino spectrum was already evaluated~\cite{zak15b}.

Another approach to the improvement of decay data for $\bar{\nu}_{e}$ summation calculations
was followed in the past by Tengblad {\it et al.}~\cite{ten89}.
They measured the spectrum of electrons emitted in the decay of individual FP
using charged particle telescopes.
Measurements were performed for up to 111 fission products 
at ISOLDE (Geneva) and OSIRIS (Studsvik). The $\beta$ spectra are converted 
into $\bar{\nu}_{e}$ spectra and both are tabulated for 95 isotopes in Ref.~\cite{rud90}.
This large set of data is free of {\it Pandemonium} and can be used both in
decay-heat and antineutrino spectrum summation calculations. 
It was pointed out by O. Bersillon during the work of WPEC-SG25~\cite{sg25}
that average $\beta$ energies from Tengblad {\it et al.}  
\cite{ten89} can be compared with average $\beta$ energies 
calculated from TAGS data obtained by Greenwood {\it et al.} \cite{gre97} for 
up to 18 fission products. The comparison shows 
that $\bar{E}_{\beta}$ energies from Tengblad {\it et al.} are systematically
larger than those from Greenwood {\it et al.}. The average difference  is $+177$~keV 
with a spread of values from $-33$~keV to $+640$~keV.
In view of the relevance of both sets of data it is important to confirm the discrepancy
and investigate possible causes.
The list of measured isotopes in  \cite{ten89,rud90} includes $^{87,88}$Br and $^{94}$Rb 
thus they can be compared with our data.

Partial results of the work presented here were already published in Ref.~\cite{tai15a}.
There we concentrated on the $\beta$-intensity distribution above $S_{n}$ and the
possible impact on $(\mathrm{n},\gamma)$ cross section estimates for unstable neutron-rich nuclei.
Here we present these results in greater detail and enhance them with the
investigation of new sources of systematic error.
In addition we present the complete $\beta$ intensity
distributions and discuss their relevance in relation to reactor decay-heat 
and anti-neutrino spectrum calculations.

\section{Measurements
\label{meas}}

\begin{figure}[h]
 \begin{center}
 \includegraphics[angle=-90, width=8.6cm]{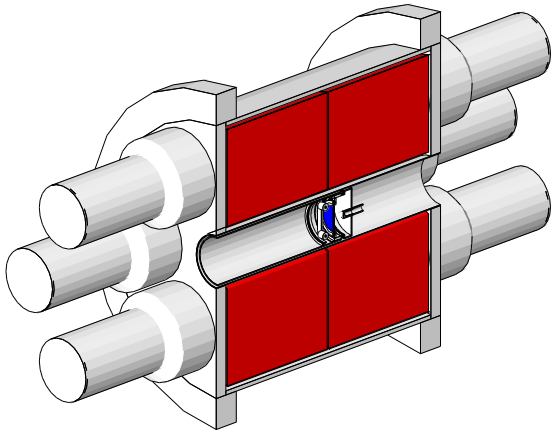}
 \caption{(Color online) Cross-sectional view of the detector geometry. BaF$_{2}$ crystals in red. Si detector in blue. 
The beam enters from the left and is deposited on the tape (not shown) in front of the Si detector.}
 \label{vtas}
 \end{center}
\end{figure}

The measurements were performed at the Cyclotron Laboratory of 
the University of Jyv\"askyl\"a. 
The isotopes of interest are produced by proton-induced fission of Uranium 
in the ion-guide source of the IGISOL Mass Separator~\cite{moo13}. 
The mass separated beam is guided to the JYFLTRAP Penning Trap~\cite{ero12}, 
for suppression of contamination. 
The JYFLTRAP mass resolving power of few tens of thousands is 
sufficient to select the isotope of interest from the rest of isobars.
The beam coming out of the trap is implanted at the centre of the spectrometer
onto a movable tape, in between two rollers holding the tape in place. 
A cross-sectional view of the detection setup is shown in Fig.~\ref{vtas}.
During the measurements the beam gate is open  
for a time period equivalent to three half-lives.
This optimizes the counting of parent decays over descendant decays.
After this period of time the tape transports the remaining activity
away and a new measuring cycle starts. 
The tape moves inside an evacuated aluminium tube of 1~mm thickness
and 47~mm diameter.
Behind the tape implantation point is placed a 0.5~mm thick Si detector 
with a diameter of 25~mm, mounted on the aluminium end-cap.
The $\beta$ detection efficiency of the Si detector is
about 30\%. The Valencia-Surrey Total Absorption Spectrometer ``Rocinante'' 
is a cylindrical 12-fold
segmented BaF$_{2}$ detector with a length and external diameter of 25~cm, 
and a longitudinal hole of 5~cm diameter. 
Each BaF$_{2}$ crystal is optically isolated by means of a thin reflector
wrapping, and viewed by a single 3" photo-multiplier tube (PMT).
The crystals are mounted inside the aluminium housing 
which has a 0.8~mm thick wall around the central hole.
The total efficiency of ``Rocinante'' for detecting a single
$\gamma$ ray with the setup described here is larger than 80\% in the energy range of interest.
The spectrometer is surrounded by 5~cm thick lead shielding to reduce
the detection of the ambient background signals.

The new spectrometer has a reduced neutron sensitivity
compared to  existing instruments based on NaI(Tl) crystals.
This is a key feature in the present measurements
as will be shown later. 
In addition, the segmentation of the detector
allows one to obtain information on $\gamma$-ray cascade 
multiplicities which helps in the data analysis.
The signal amplitudes from the 12 independent PMTs
are digitized in a peak sensing analog-to-digital converter (ADC)
and stored on disk for each event.
The event trigger is provided whenever the hardware sum of the PMT signals
fires a constant fraction discriminator (CFD).
The signal from the Si detector is processed in an analogous manner
providing another trigger for read-out and storage of events.
In the off-line analysis the PMT signals 
are gain matched and those surpassing a common threshold of 65~keV are added to obtain the
total absorption spectrum. 
The gain-matching procedure
uses as a reference the position of the $\alpha$-peaks visible in the energy spectra
coming from the Ra contamination always present in BaF$_2$ crystals. 
In order to eliminate this intrinsic background 
as well as the ambient background
we use in the present analysis $\beta$-gated total absorption spectra.
The threshold in the Si $\beta$ detector is set to 100~keV.
Nevertheless other sources of background
need to be taken into account. 

%%%%%  Figure 1  %%%%%

Firstly there is the decay descendant contamination,
which was computed using Monte Carlo (MC) simulations performed
with the Geant4 simulation toolkit~\cite{geant4}. 
In the case of daughter decay we use an event generator based on  
$\beta$ intensity distributions and $\gamma$ branching ratios
obtained from the well known~\cite{br87,br88,rb94} decay scheme. 
The normalization of the daughter contamination  
is estimated from the known half-lives and the
measurement cycle time information and eventually adjusted
to provide the best fit to the recorded spectrum. 
The measurement of $^{88}$Br was accidentally
contaminated with $^{94}$Y, the long-lived grand-daughter of $^{94}$Rb that was measured
immediately beforehand. It was treated in the same manner.

The contamination due to the $\beta$-delayed neutron branch
is more challenging. The decay simulation must explicitly include the neutrons emitted.
These neutrons interact with detector materials producing $\gamma$-rays through inelastic
and capture processes, which are readily detected in the spectrometer.
A specific event generator was implemented which reproduces the known
neutron energy distribution, taken from~\cite{endfb71}, and the known $\gamma$-ray
intensity in the final nucleus, taken from~\cite{br87,br88,rb94}. The event generator 
requires the $\beta$ intensity distribution followed by
neutron emission $I_{\beta n} (E_{x})$ and the branching to each level
in the final nucleus. We have used the simplifying
assumption that this branching is independent of the excitation energy 
in the daughter nucleus to obtain $I_{\beta n} (E_{x})$ from the neutron spectrum. 
The deconvolution of the neutron spectrum using calculated $\Gamma_{\gamma}$ and $\Gamma_{n}$
within the Hauser-Feshbach model
(see below) leads to similar $I_{\beta n} (E_{x})$ but does not 
reproduce the $\gamma$ intensity in the final
nucleus.
Another issue is whether the interaction
of neutrons with the detector can be simulated accurately. 
We have shown recently~\cite{tai15b} this to be the case
for a LaBr$_{3}$:Ce detector,
provided that Geant4 is updated  
with the newest neutron data libraries and the original capture cascade 
generator is substituted by an improved one
based on the nuclear statistical model.
We have followed the same approach for our BaF$_{2}$ detector.
The normalization factor of the $\beta$-delayed neutron decay 
contamination is fixed by the $P_{n}$ value.

Another important source of spectrum distortion is the
summing-pileup of events. 
If more than one event arrives within the same ADC event gate, a signal with 
the wrong energy will be stored in the spectrum.
Apart from the electronic pulse pile-up effect for a single crystal,
which can be calculated using the methodology described in~\cite{can99a},
one must 
consider the summing of signals from different crystals. 
A new Monte Carlo procedure to calculate their combined contribution has been developed. 
The procedure is based on the superposition of two recorded events, selected randomly. 
The time of arrival of the second event is sampled randomly within the ADC gate length. 
The normalization of the resulting summing-pileup spectrum is fixed by the 
true rate and the ADC gate length~\cite{can99a}. 
To calculate the rate
a dead time correction is necessary and this is obtained by counting
the signals from a fixed frequency pulse generator feeding the preamplifier. 
The use of real events to calculate the
spectrum distortion is valid if the actual summing-pileup rate is small enough.
For this reason we kept the overall rate during the measurements below 7~kcps.
The method is
validated with measurements of laboratory sources.

%%%%%  Figure 2  %%%%%
\begin{figure}[h]
 \begin{center}
 \includegraphics[width=8.6cm]{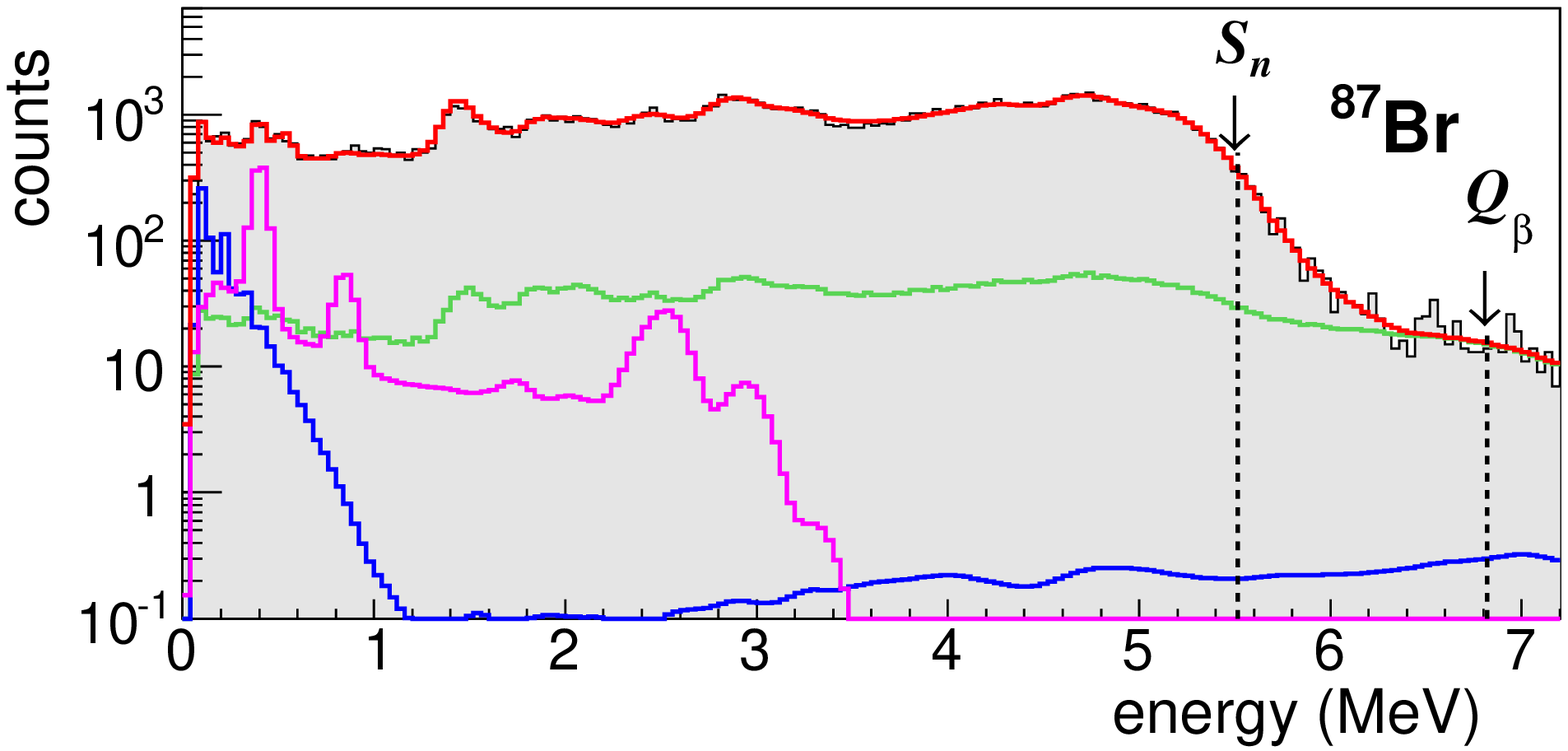}
 \includegraphics[width=8.6cm]{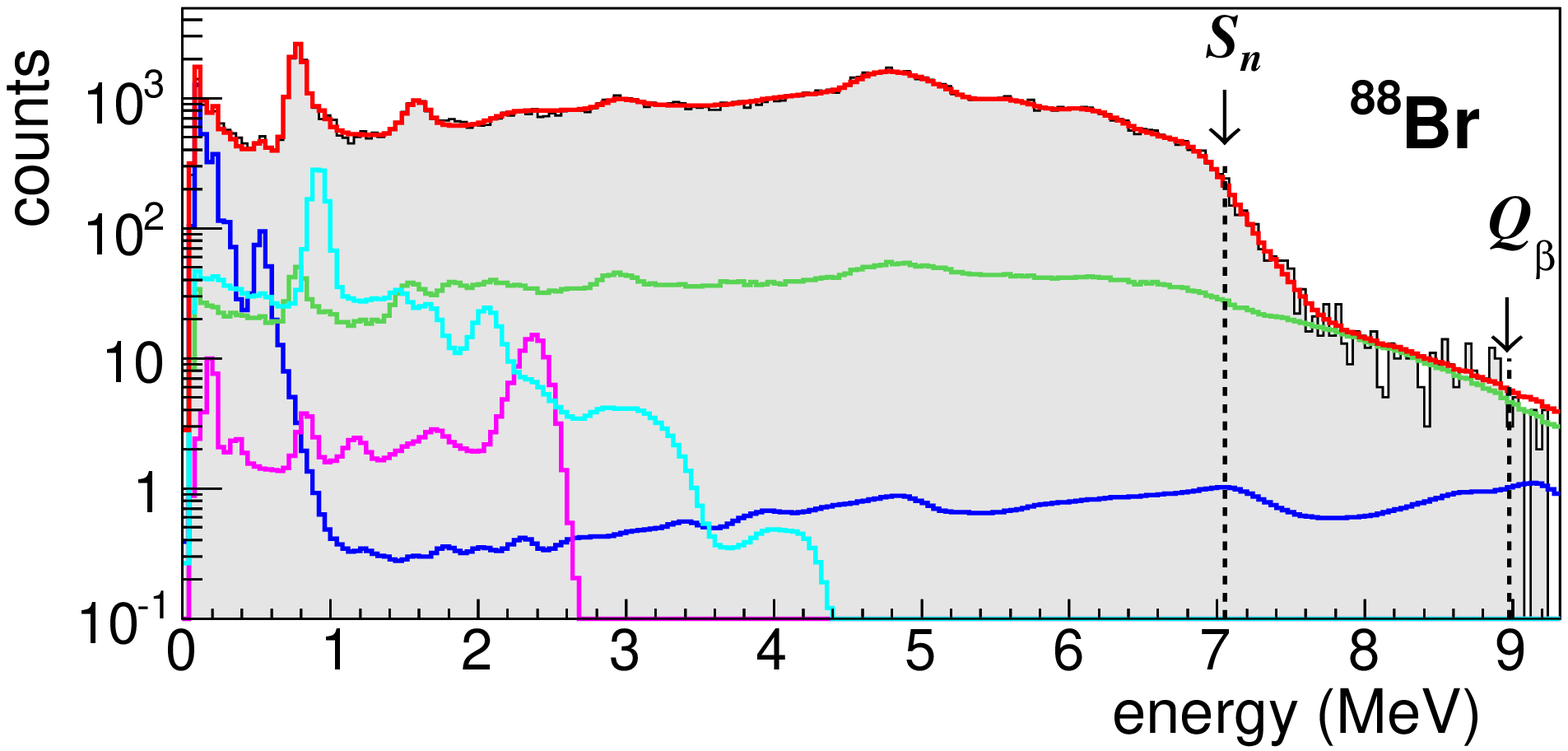}
 \includegraphics[width=8.6cm]{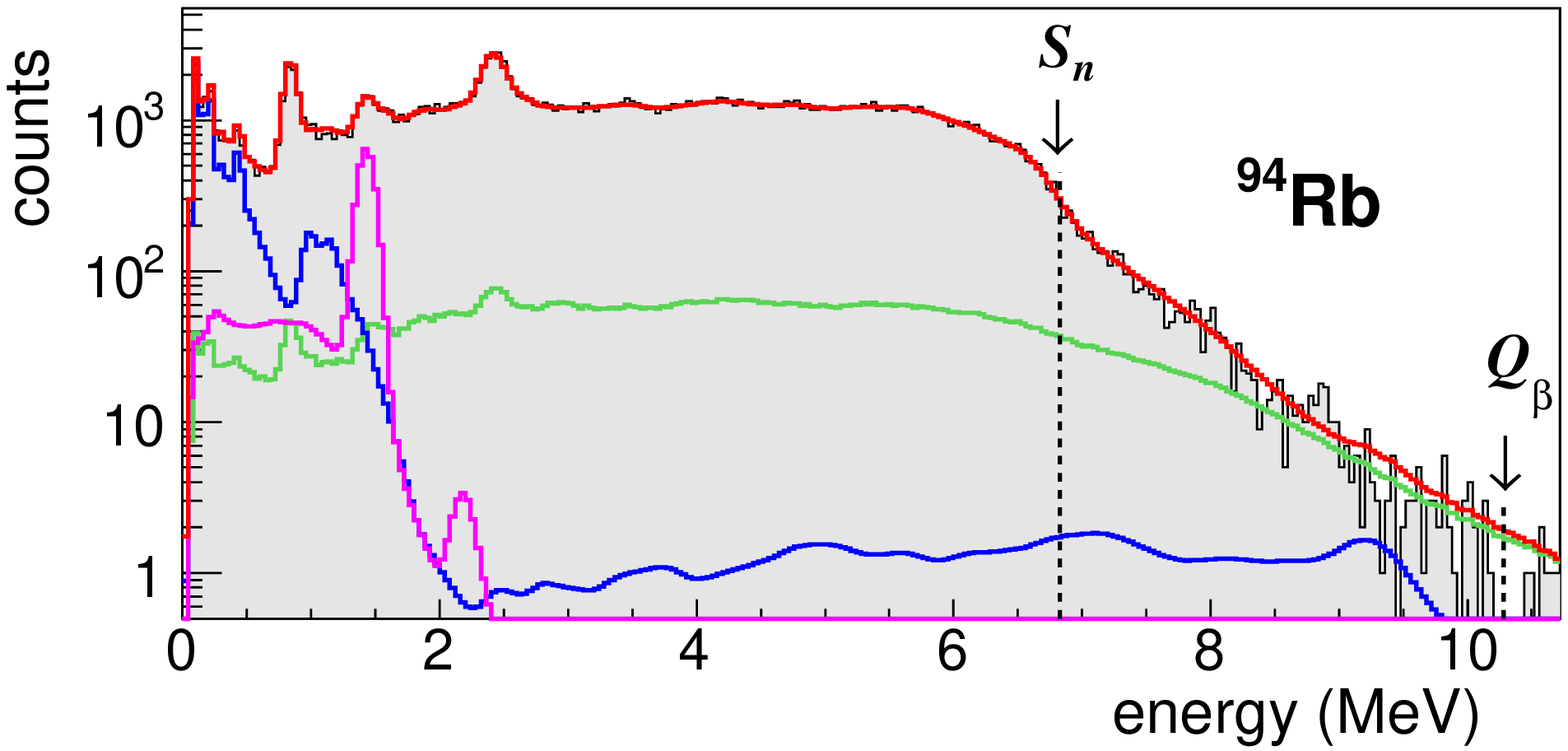}
 \caption{(Color online)  Relevant histograms for the analysis: parent decay
(gray filled), daughter decay (pink), delayed neutron decay (dark blue),  
accidental contamination (light blue), summing-pileup contribution (green), 
reconstructed spectrum (red). 
See text for details. The neutron separation energy $S_{n}$ and 
decay energy window $Q_{\beta}$ are also indicated.}
 \label{tasspec}
 \end{center}
\end{figure}

Several sources, $^{22}$Na, $^{24}$Na, $^{60}$Co and $^{137}$Cs, were used
to determine both the energy calibration and the resolution versus energy
dependency of the spectrometer. The latter is needed to widen the MC
simulated response and is parametrized in the form 
of a Gaussian with $\sigma_{E} = \sqrt{a E +b E^{2}}$. 
The highest calibration point is at 4.123~MeV. 
At this energy the energy resolution (FWHM) is 265~keV which becomes 455~keV at 10~MeV.
The ungated spectra measured with the sources
serve also to verify the accuracy of the Geant4 MC simulations of the
spectrometer response to the decay. 
This requires a detailed description in the simulation code of all
materials in the measurement setup (tape transport system and detectors) . 
The use of $\beta$-gated spectra in the analysis requires additional verifications
of the simulation. Due to the existence of an electronic threshold in the Si detector (100~keV)
and the continuum nature of the $\beta$ spectrum the efficiency for $\beta$-detection
has a strong dependency with endpoint energy up to about 2~MeV.
It should be noted that this affects the spectral region 
above $S_{n}$ in which we are particularly interested.
To investigate  whether the MC simulation can reproduce this
energy dependency accurately we used the information
from a separate experiment~\cite{agr16} measuring $P_{n}$ values with the
neutron counter BELEN and
the same $\beta$ detector and implantation setup. Several
$\beta$-delayed neutron emitters with known neutron energy
spectra were measured, including $^{88}$Br,
$^{94,95}$Rb and $^{137}$I. 
They have different neutron emission windows
$Q_{\beta}- S_{n}$, therefore the neutron-gated $\beta$ efficiency  
samples different portions of the low energy part of the efficiency curve.
Indeed the measured average $\beta$ detection efficiency for each isotope
changes by as much as 25\%.
Using the above mentioned $\beta$-delayed neutron decay generator
in Geant4 we 
are able to reproduce the isotope dependent efficiency to
within better than 4\%, determining the level of accuracy of the simulation.

Figure~\ref{tasspec} shows the $\beta$-gated TAGS spectrum measured 
for all three isotopes.
Also shown is the contribution to the measured spectra 
of the daughter decay, the neutron decay branch,
and the summing-pileup effect.
In the case of $^{88}$Br it also includes 
the contribution of the accidental contamination with $^{94}$Y decay.  
Note that there are net counts above the background beyond the
neutron separation energy. The fraction of counts that are to
be attributed to states above $S_{n}$ populated in the decay
de-exciting by $\gamma$-ray emission
is obtained after deconvolution with the spectrometer response. 
In this region
the major background contribution comes from summing-pileup which is well reproduced
by the calculation as can be observed. 
The contribution of neutron capture $\gamma$-rays in the detector materials
is much smaller, thanks to the low neutron sensitivity of BaF$_{2}$,
as can be seen. The contribution of $\gamma$-rays coming from neutron inelastic scattering
is important at energies below 1~MeV. 

\section{Analysis
\label{anal}}

The analysis of the $\beta$-gated spectra follows the method developed by the Valencia 
group~\cite{tai07a,tai07b}. 
The deconvolution of spectra with the spectrometer response to the
decay is performed using the Expectation-Maximization (EM) 
algorithm described there.
The spectrometer response is constructed in two steps. First the response to
electromagnetic
cascades is calculated from a set of branching ratios and the MC calculated 
response to individual $\gamma$-rays.
In the simulation we use a single crystal low energy threshold of 65~keV from experiment.
When necessary, the electron conversion process is taken into account while building
the response~\cite{can99b}. 
Branching ratios are taken from~\cite{br87,br88,rb94} 
for the low energy part  of the level scheme.
In the present case this involves 4 levels up to 1.6~MeV for $^{87}$Kr, 
8 levels up to 2.5~MeV for $^{88}$Kr 
and 11 levels up to 2.8~MeV for $^{94}$Sr.
The excitation energy range above the last discrete level is treated as a
continuum and is
divided into 40~keV bins. Average 
branching ratios for each bin are calculated from the NLD and PSF 
as prescribed by the nuclear statistical model. 
We use the NLD calculated using a Hartree-Fock-Bogoliubov (HFB) plus combinatorial
approach adjusted to experimental information~\cite{gor08,ripl3},
which includes parity dependence.
%(number of levels at low energy and
%number of resonances at high excitation energies when available). 
The PSF is obtained 
from Generalized Lorentzian (E1 transitions) or Lorentzian (M1 and E2 transitions) 
parametrization using the parameters 
recommended in the RIPL-3 reference input parameter library~\cite{ripl3}. In the second
step of the response construction, the previously obtained electromagnetic 
response for each level or energy bin is convoluted with the simulated
response to a $\beta$ continuum of allowed shape.
The $\beta$ response is obtained under the condition that the energy deposited in the Si detector 
is above the 100 keV threshold.   

The spins and parities of some of the discrete states 
in the daughter nucleus are ambiguous but they are needed in order
to calculate the branching ratio from states in the continuum.
In the analysis different spin-parity values are tested  and those giving the best fit
to the spectrum are taken. 
The spin and parity of the parent nucleus ground state is also uncertain,
however it determines the spin and parity of the states
populated in the continuum needed to construct the branching ratio matrix.
We assume that the Gamow-Teller selection rule applies for decays into the continuum, 
i.e., the parity does not
change and the spin change fulfill $| \Delta J | \leq 1$.
In the calculation of the branching ratios we further assume that
different spins $J$ are populated according to the spin statistical weight
$2J+1$.
Our choices of spin and parity for the ground state 
are $3/2^{-}$ for $^{87}$Br, $1^{-}$ for $^{88}$Br and $3^{-}$ for
$^{94}$Rb, based again on the quality of reproduction of
the measured spectra.
The spin-parity of $^{87}$Br is given as $3/2^{-}$ in Ref.~\cite{br87},
however Ref.~\cite{por06} proposes $5/2^{-}$.
We do not find significant differences in the analysis 
assuming these two values and we choose the former.
The spin-parity of $^{88}$Br is uncertain and is given as $(2^{-})$ in Ref.~\cite{br88}.
However Ref.~\cite{gen99} suggests $1^{-}$. In our analysis
we use the latter value since it clearly provides a much better reproduction of
the measured TAGS spectrum. In the case of $^{94}$Rb $3(^{-})$ is proposed~\cite{rb94}
and is adopted, since other alternatives 
did not lead to a better reproduction of the spectrum.

In the analysis we permit decays to all discrete states,
many of which are of the forbidden type. Forbidden transitions to the ground state or
low lying excited states are known to occur in this region of the nuclear chart.
Indeed sizable decay intensity for some forbidden transitions is obtained in our analysis.
In the case of $^{87}$Br we find a ground state intensity $I^{gs}_{\beta}=10.1\%$ quite
close to 12\%, the quoted value in Ref.~\cite{br87}. However in contrast to \cite{br87}, the first four excited states
included in the discrete part receive negligible intensity. The summed decay intensity to the discrete
part becomes 51\% of  that in Ref.~\cite{br87}. In Ref.~\cite{br88} an upper limit of 11\%
is given for the $^{88}$Br ground state decay intensity, 
and a sizable intensity is quoted for some of the eight excited
states included in the analysis. We obtain 4.7\% and 5.6\% for the $\beta$ intensity 
to the ground state and first excited state
respectively, and small or negligible intensity for the remaining states. Overall the intensity
to this part of the level scheme is reduced by 64\%. 
No intensity is assigned in \cite{rb94} to $^{94}$Rb 
decaying to the ground state (third forbidden)
and first excited state (first forbidden). In our analysis
we forbid the decay to those states after verifying that the decay intensity  
obtained when left free
is only 0.5\% and 0.02\% respectively.
A large decay intensity of 23.7\%
is observed for the allowed transition to the state at $E_{x}=2414$~keV, even larger than the value of
21.4\% found in \cite{rb94}. The intensity to the discrete level scheme included in our analysis (11 states)
is 78\% of that in ENSDF.

In the final analysis we applied a correction to branching ratios
deduced from the statistical model.
The aim is to obtain a spectrometer response that is as realistic as possible.
We scale the calculated branching ratios going from the unknown part of the level scheme 
to discrete levels in the known part of the level scheme, 
in order to reproduce the observed $\gamma$-ray intensities
as tabulated in Ref.~\cite{br87,br88,rb94}. 
Here we are making the assumption that the absolute $\gamma$ intensity is correctly
determined in the high resolution measurements for the lowest excited levels.
We found that this adjustment did not lead to significant changes in the quality of reproduction of the
measured TAGS spectra and has a small impact on the results of the deconvolution.

%%%%%  Figure 3  %%%%%
\begin{figure}[h]
 \begin{center}
 \includegraphics[width=8.6cm]{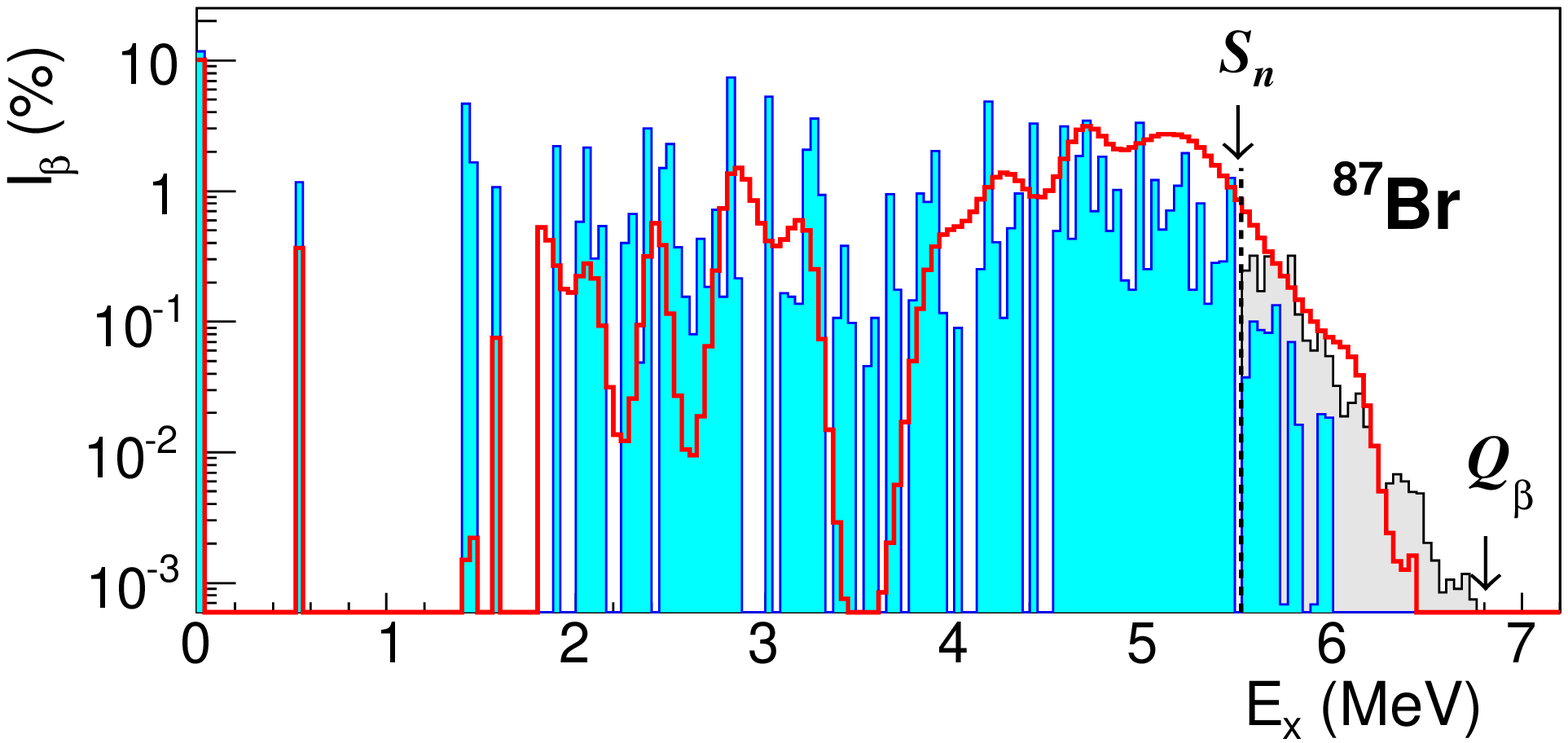}
 \includegraphics[width=8.6cm]{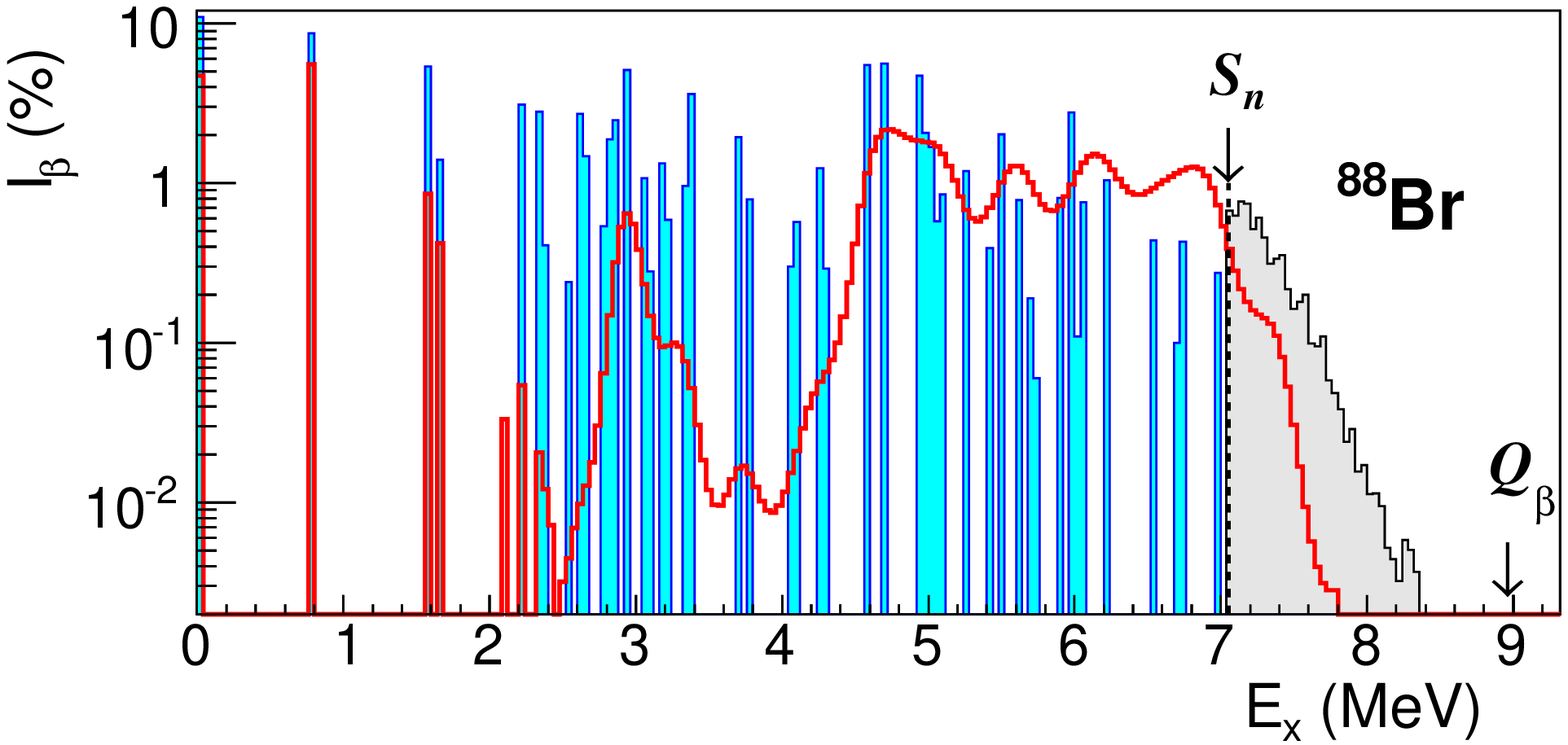}
 \includegraphics[width=8.6cm]{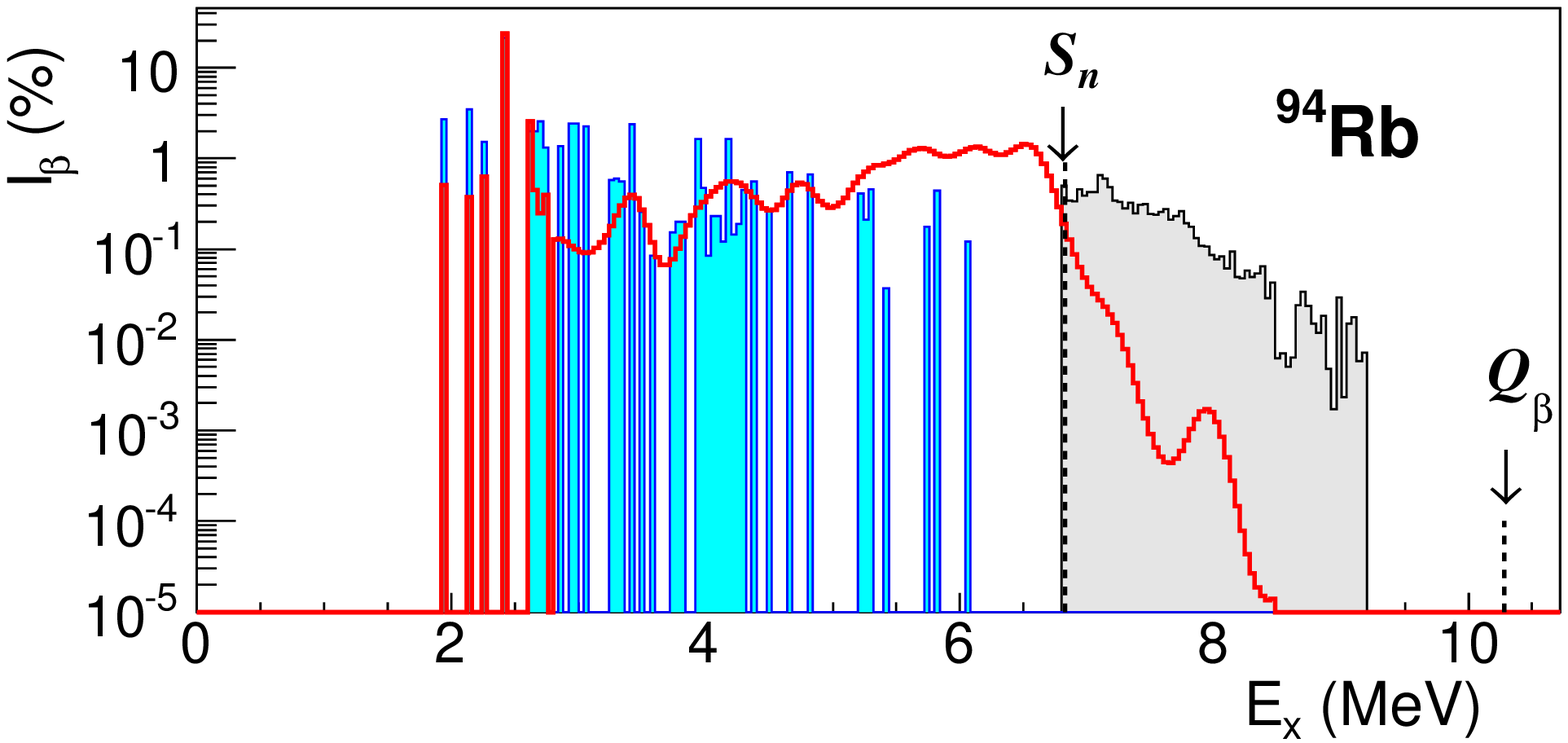}
 \caption{(Color online)  Beta intensity distributions: TAGS result (red line), 
high-resolution measurements (blue filled),  from delayed neutron spectrum (gray filled).
See text for details.}
 \label{tasint}
 \end{center}
\end{figure}

Figure~\ref{tasint} shows 
the final $\beta$ intensity distribution $I_{\beta \gamma} (E_{x})$ resulting from the deconvolution 
of TAGS spectra for all three isotopes with the chosen branching ratio matrices. 
The intensity is normalized to $(100-P_{n})$\%.
In each case the spectrum reconstructed with this intensity
distribution gives a good reproduction of the measured spectrum as can be
seen in Fig.~\ref{tasspec}. 
The full $\beta$ intensity distribution including statistical uncertainties
is given as Supplemental Material to this article~\cite{supplement}.
The uncertainty due to the statistics in the data is computed 
according the prescription given in Ref.~\cite{tai07b} and is 
very small. 

We evaluate the impact of several sources of systematic uncertainty on the shape 
of the $\beta$ intensity distribution. These include both uncertainties in the 
calculated decay response and uncertainties in the subtraction of background components.
To study their effect we follow a similar procedure in all the cases. 
The chosen systematic parameter is
varied and a new deconvolution is performed until we observe an appreciable
deterioration in the reproduction of the measured spectrum.
This is quantified by the increase of chi-square between
the measured and reconstructed spectra.
In this way we obtain the maximum acceptable deviation of the
$I_{\beta \gamma} (E_{x})$ from the adopted solution for each
investigated systematic uncertainty.
As a reference the maximum chi-square increase is always below 5\%.

Uncertainties in the calculated decay response are of two types. 
Uncertainties in the branching ratio matrix, which were discussed above,
and uncertainties in the MC simulation of the response to $\gamma$ 
and $\beta$ radiation. As already explained we take great care
to describe accurately the geometry used in the Geant4 simulation, 
which is validated from the comparison with measurements with laboratory
sources. However these sources emit $\beta$ particles with rather small
energy and they are not useful to verify the $\beta$ response.
The simulated $\beta$ efficiency of the Si detector and in particular its
variation with endpoint energy was studied in a separate measurement~\cite{agr16}
as already discussed. The response of the spectrometer to $\beta$ particles
depositing energy in the Si is not easy to verify.
The response is a mixture of $\beta$ penetration and
secondary radiation produced in dead materials.
The accurate simulation of the interaction of low energy electrons
is a challenging task
for any MC code. They rely on models to describe
the slowing down of electrons and changes
in their trajectory. Typically a number of tracking parameters
are tuned to obtain reliable results.
We use in the present simulations the
{\it Livermore Electromagnetic Physics List} of
Geant4 (version 9.2.p2) with original tracking parameters.
This physics list has been developed for high
accuracy tracking of low energy particles.
We verified that limiting the tracking step length (parameter {\it StepMax})
to values much smaller than default values,
increased computing time considerably but did not significantly affect the simulated response.
To study the effect of a possible systematic error on the $\beta$ response 
we take a pragmatic approach.
We scale arbitrarily the simulated spectrometer response
while keeping the same $\beta$ efficiency. 
In this way we find that solutions corresponding to 
changes of $\pm10$\% in the $\beta$ response normalization
represent the maximum deviation with respect to the adopted
solution that can be accepted.

%%%%%  Figure 4  %%%%%
\begin{figure}[h]
 \begin{center}
 \includegraphics[width=8.6cm]{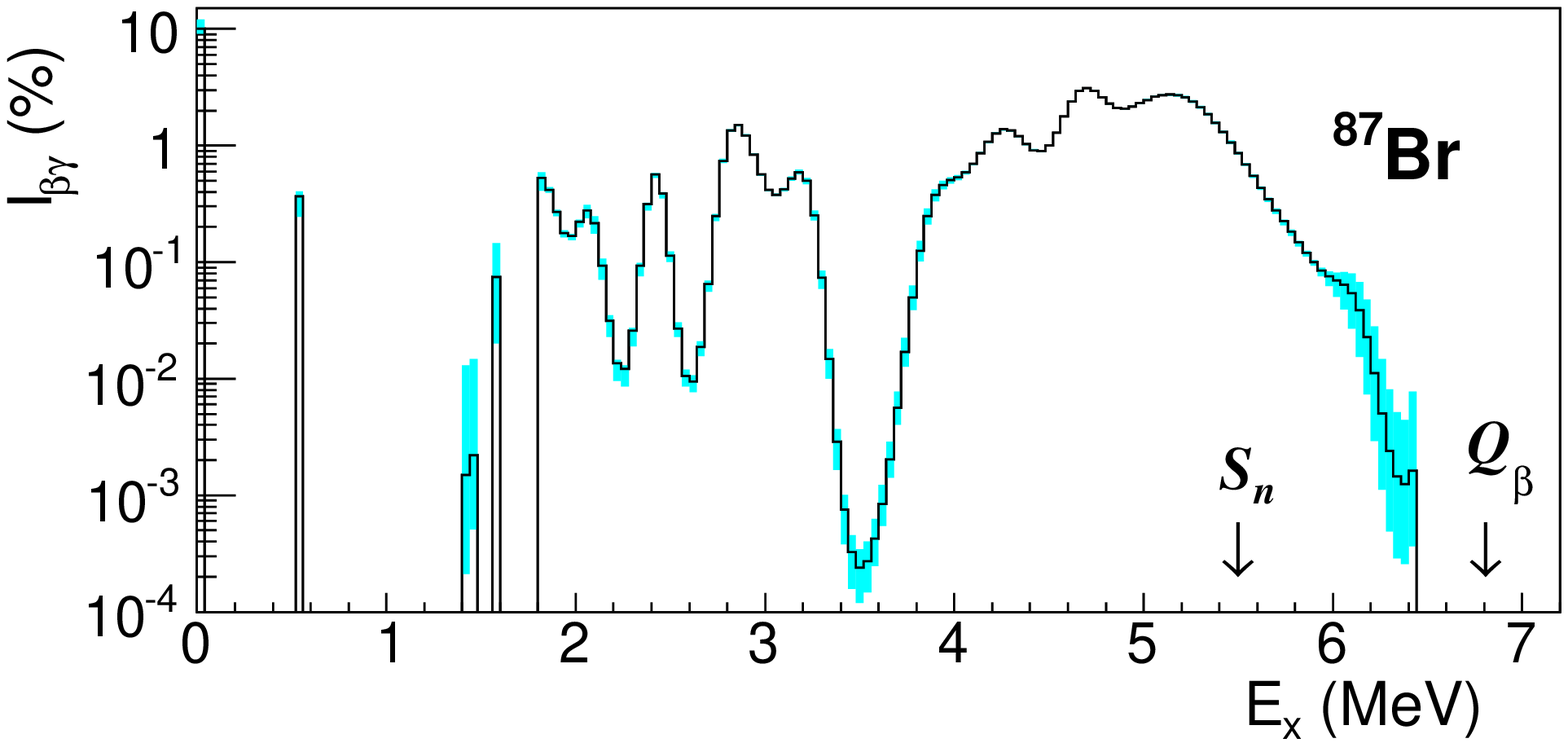}
 \includegraphics[width=8.6cm]{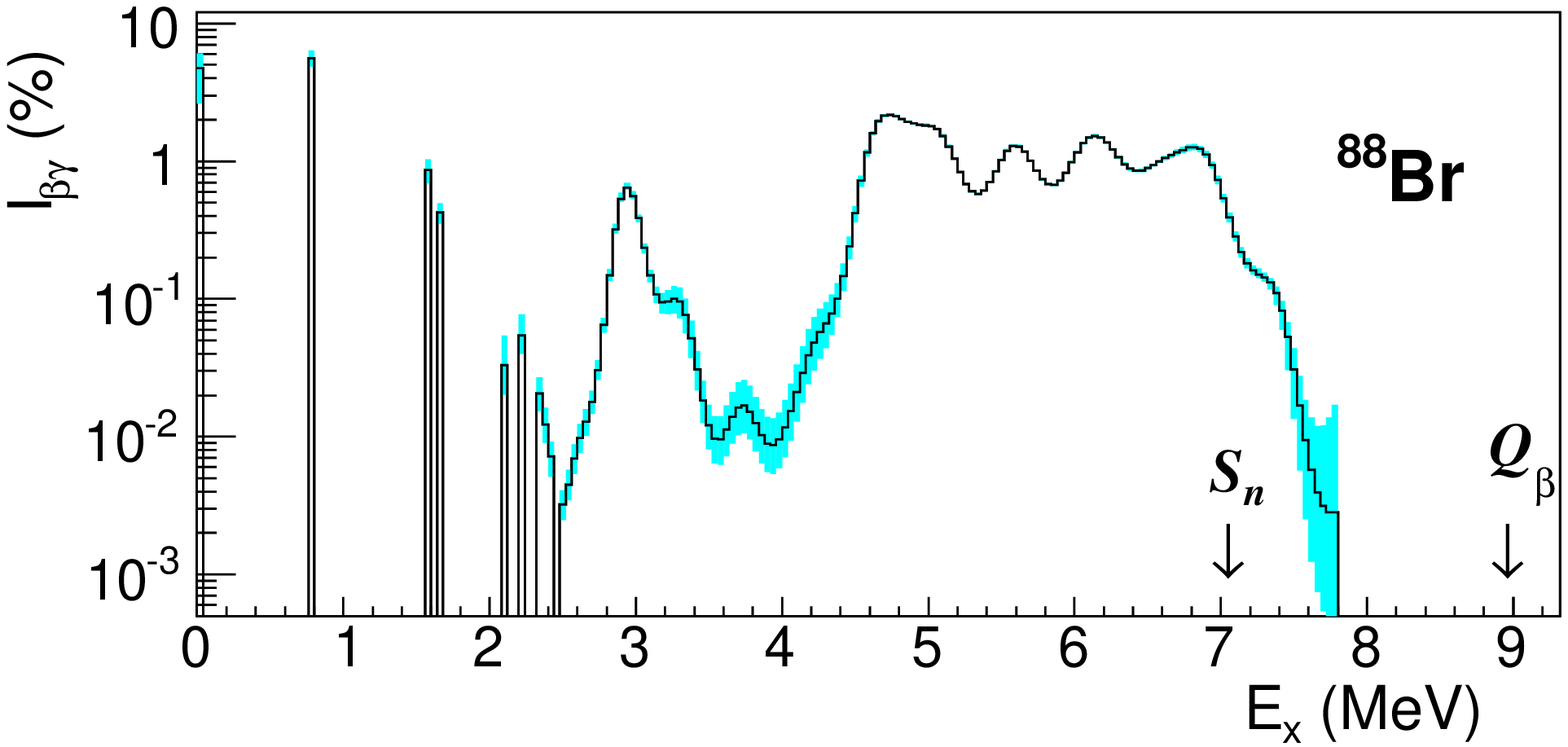}
 \includegraphics[width=8.6cm]{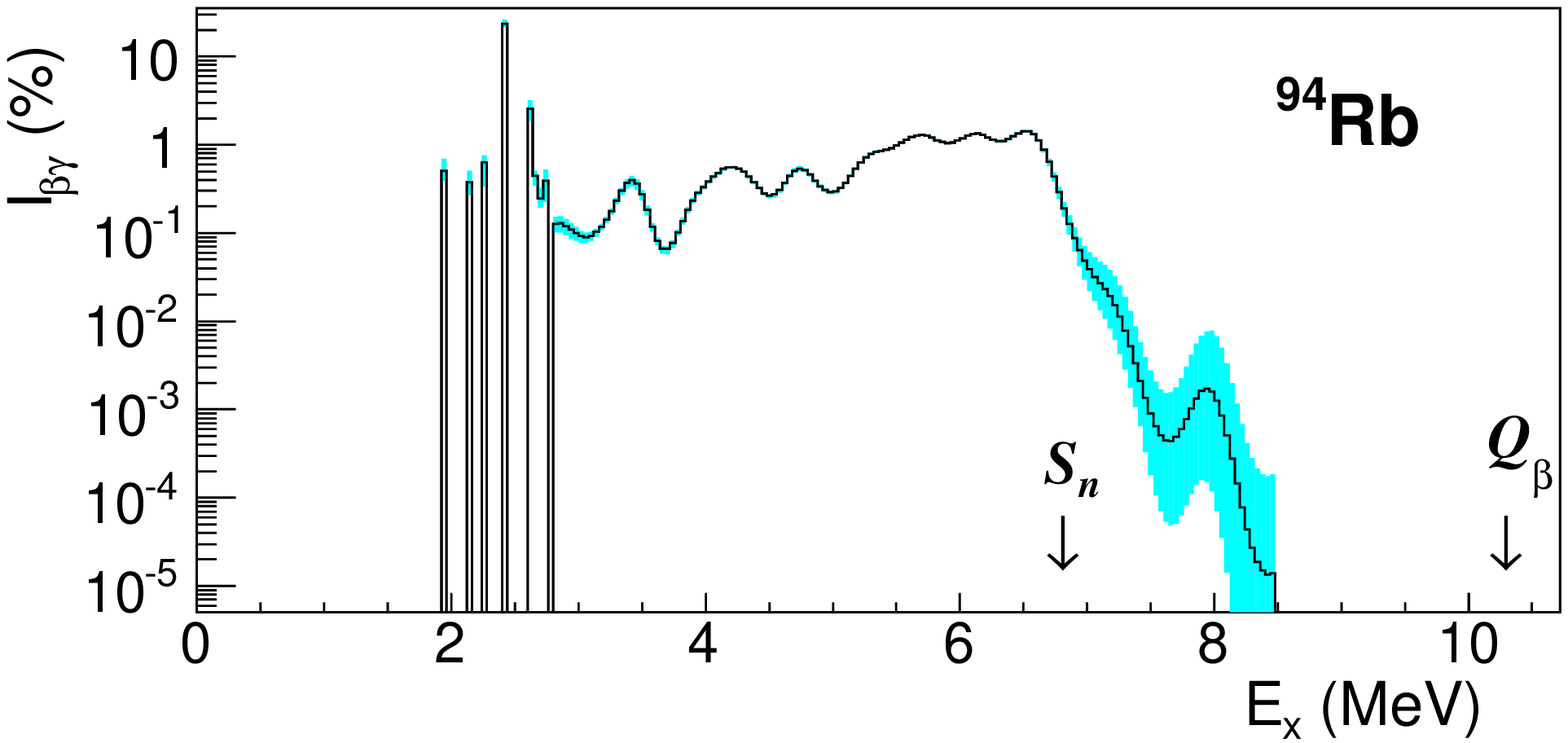}
 \caption{(Color online) Beta intensity distributions from TAGS. The thin black line is the adopted solution, 
the light blue filled region indicates the spread of solutions due to the systematic effects investigated.
See text for details.}
 \label{fedmaxerr}
 \end{center}
\end{figure}

The individual $\gamma$ response is well tested up to $E_{\gamma}=2.754$~MeV, the maximum
energy for the $^{24}$Na source. To investigate the effect
of a possible systematic error in the total $\gamma$ efficiency  $\varepsilon_{\gamma}$ or in the
peak-to-total ratio (P/T) we introduce a model that varies linearly
one of the two parameters, $\varepsilon_{\gamma}$ or P/T, above $E_{\gamma} = 3$~MeV.
We found that variations of $\varepsilon_{\gamma}$ amounting to $\pm15$\% at 
 $E_{\gamma} = 10$~MeV or variations of P/T amounting to $\pm30$\% at
the same energy are the maximum allowed by good reproduction of the spectrum.
When considering these numbers one should bear in mind that the de-excitation 
of highly excited states populated in the decay of the three isotopes proceeds 
with an average $\gamma$ multiplicity of 2 to 4 in such a way that the energy of most 
$\gamma$ rays in the decay does not exceed 3~MeV.

Uncertainties in the normalization of background components also have an impact
on the $\beta$ intensity distribution. We consider the two main components,
summing-pileup which affects the high energy part of the spectrum, and the $\beta$-delayed
neutron decay branch, which affects the low energy part of the spectrum
(see Fig.~\ref{tasspec}).
The component due to summing-pileup is normalized
using the same ADC gate length (5~$\mu$s) for all three isotopes. 
We estimate however that the reproduction of the end-part of the spectra 
allows for a variation of up to $\pm15$\% in the normalization factor. 
The normalization of the $\beta$-delayed neutron decay component is fixed by the $P_{n}$ value.
Likewise we find that the reproduction of the low energy part of the spectrum 
allows for a variation of up to $\pm15$\% in the normalization factor.

Finally we also check the impact on the result associated with the use of a different 
deconvolution algorithm, by using the Maximum Entropy Method as described in Ref.~\cite{tai07b}.
This leads to changes in the $I_{\beta} (E_{x})$ noticeable both at the high-energy end
and at low $E_{x}$.

There is no straightforward way to quantify and combine the systematic uncertainties 
associated with the effects investigated. One of the reasons is that they
are not independent since we are requiring reproduction of the data.
It would have been a formidable task to explore in a correlated way the full parameter
space. We use a different point of view here. The solutions we obtain
through the systematic variation of each parameter represent maximum
deviations from the adopted solution, thus altogether define an estimate of the
space of solutions compatible with the data. This is represented in a graphical
way in Fig.~\ref{fedmaxerr} showing the envelope of the different solutions
described above corresponding to the maximum accepted deviation from the adopted solution.
In total there are 14 solutions for $^{87}$Br, 13 for $^{88}$Br, and  15 for $^{94}$Rb.
As can be seen the different solutions differ little except for specific $E_{x}$ regions,
where the $\beta$ intensity is low, in particular at the high energy end of the distribution
or at low excitation energy.

\section{Average beta and gamma decay energies and decay heat
\label{heat}}

\begin{figure}[h]
 \begin{center}
 \includegraphics[width=8.6cm]{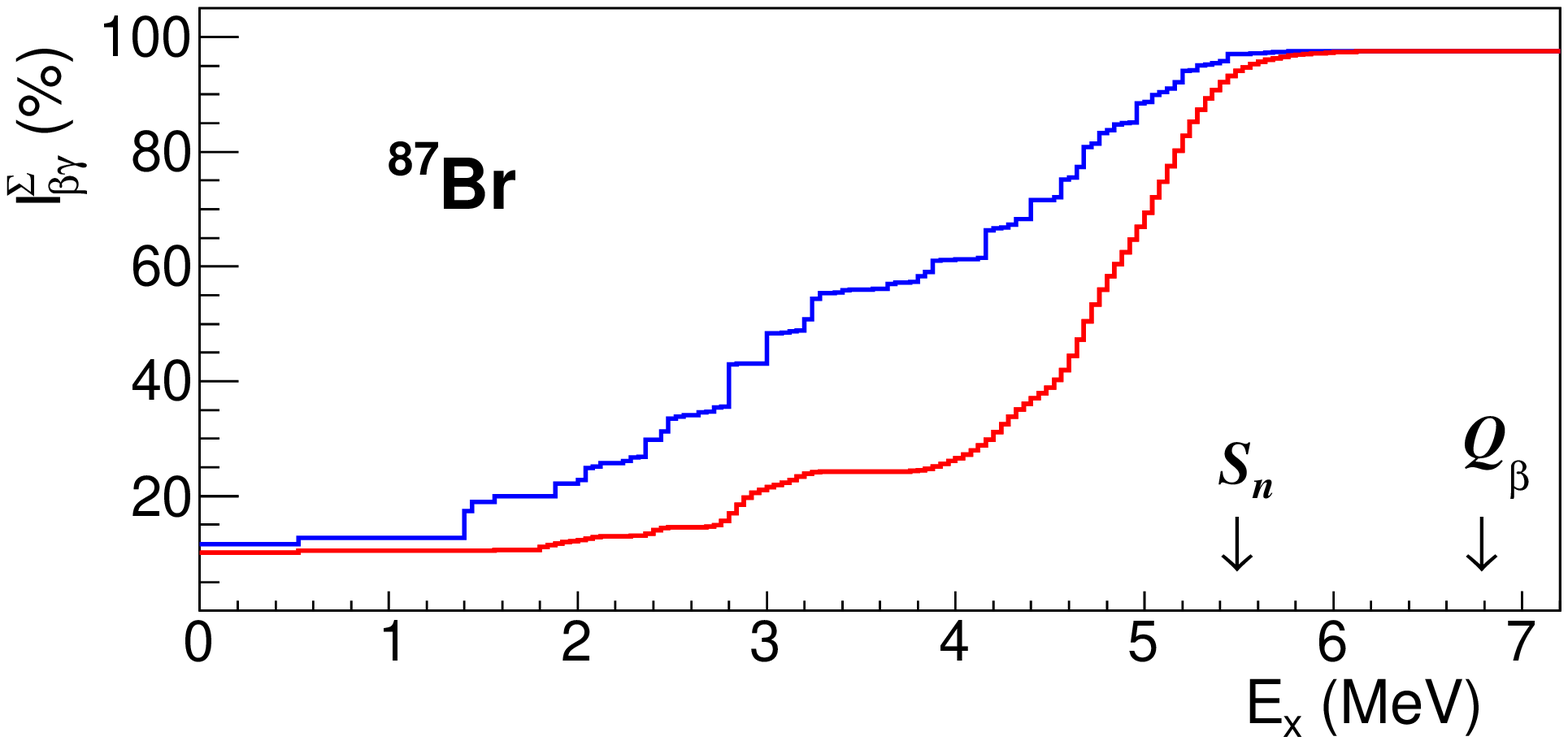}
 \includegraphics[width=8.6cm]{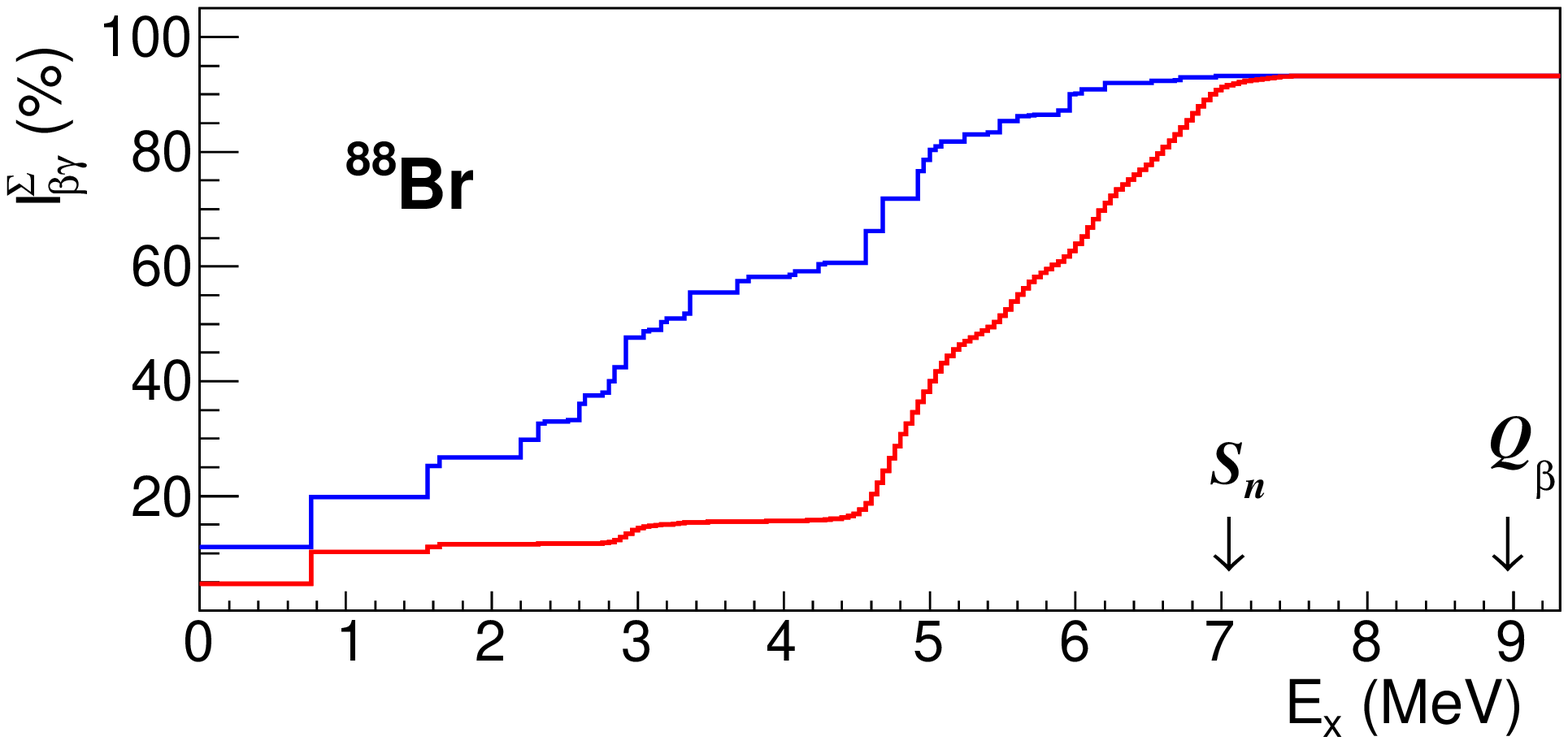}
 \includegraphics[width=8.6cm]{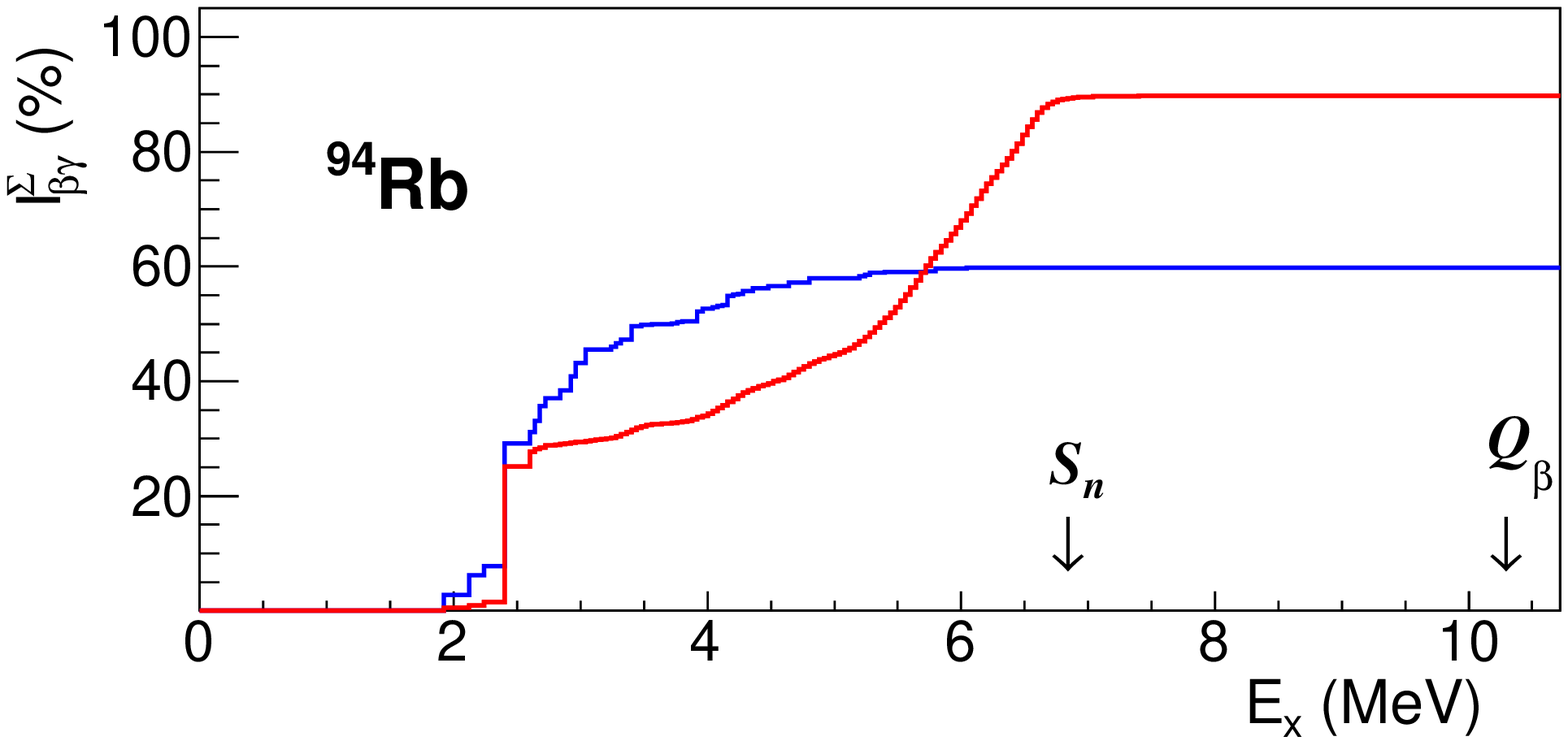}
 \caption{(Color online)  Accumulated $\beta$ intensity distribution $I^{\Sigma}_{\beta \gamma}$: 
TAGS result (red line), high-resolution measurements (blue line). }
 \label{acctasint}
 \end{center}
\end{figure}

Figure~\ref{tasint} shows in addition to $I_{\beta \gamma} (E_{x})$ 
obtained from our TAGS data 
the intensity obtained from high-resolution
measurements retrieved from the ENSDF data base~\cite{br87,br88,rb94}. 
The effect of {\it Pandemonium} is visible here.
Our results show a redistribution of $I_{\beta \gamma}(E_{x})$ towards high $E_{x}$,
which is significant for  $^{87}$Br,
and very large for $^{88}$Br and $^{94}$Rb. 
This is even clearer in the accumulated $\beta$ intensity distribution  
as a function of excitation energy 
$I^{\Sigma}_{\beta \gamma} (E_{x}) = \int_{0}^{E_{x}} I_{\beta \gamma} (E) dE$, 
depicted in Fig.~\ref{acctasint}.
The intensity is normalized to 100\%$-P_{n}$ except in the case of the $^{94}$Rb
ENSDF intensity  that only reaches 
59.8\% since the evaluators of Ref.~\cite{rb94} recognize the incompleteness 
of the decay scheme.

%%%%%  Figure 5  %%%%%

Table~\ref{averene} shows $\bar{E}_{\gamma}$ and $\bar{E}_{\beta}$
obtained from $I_{\beta \gamma} (E_{x})$ 
using Eq.~\ref{eq:avereneg} and Eq.~\ref{eq:avereneb} respectively.
The $\beta$ continuum and its average energy $\langle E_{\beta} (Q_{\beta}-E_{x}) \rangle$ 
for each $E_{x}$
is calculated using subroutines extracted from the LOGFT program package
maintained by NNDC (Brookhaven)~\cite{logft}.
In the calculations we assume an allowed $\beta$ shape.
As can be seen in Table~\ref{averene} the redistribution of $\beta$ intensity leads to large 
differences in the average emission
energies when comparing high resolution data (ENSDF)
and the present TAGS data. 
The difference has opposite directions for $\gamma$ and $\beta$ energies, as expected,
except in the case of $^{94}$Rb due to the use of a different normalization.
For $\bar{E}_{\gamma}$ the difference is 0.9~MeV for $^{87}$Br, 1.7~MeV for $^{88}$Br,
and 2.3~MeV for $^{94}$Rb.
The uncertainty quoted on the TAGS average energies in Table~\ref{averene} is systematic 
since the contribution of statistical uncertainties in the case of $I_{\beta\gamma} (E_{x})$ is negligible.
The values of  $\bar{E}_{\gamma}$ and $\bar{E}_{\beta}$ were computed
for each intensity distribution that was used to define the space of accepted solutions
in Fig.~\ref{fedmaxerr}, and the maximum positive and negative difference 
with respect to the adopted solution is the value quoted in the Table.

%%%%%  Table 2  %%%%%
\begin{table}[h]
\caption{Average $\gamma$ and $\beta$ energies calculated using $I_{\beta \gamma} (E_{x})$ intensity
distributions from ENSDF~\cite{br87,br88,rb94} and present TAGS data. The contribution of the 
$\beta$-delayed neutron branch is not included.} 

\begin{center}
%\resizebox{5.7cm}{!}{
\begin{tabular}{ccccc} \hline \hline
 & \multicolumn{2}{c}{$\bar{E}_{\gamma} \mathrm{(keV)}$} & \multicolumn{2}{c}{$\bar{E}_{\beta} \mathrm{(keV)}$} \\ \hline
Isotope& ENSDF & TAGS & ENSDF & TAGS \\ \hline
$^{87}$Br & 3009 & $3938^{+40}_{-67}$ & 1599 & $1159^{+32}_{-19}$  \\
$^{88}$Br & 2892 & $4609^{+78}_{-67}$ & 2491 & $1665^{+32}_{-38}$  \\
$^{94}$Rb & 1729 & $4063^{+62}_{-66}$ & 2019 & $2329^{+32}_{-30}$ \\ \hline \hline
\end{tabular}
%}
\end{center}
\label{averene}
\end{table}

\begin{figure}[h]
 \begin{center}
 \includegraphics[width=8.6cm]{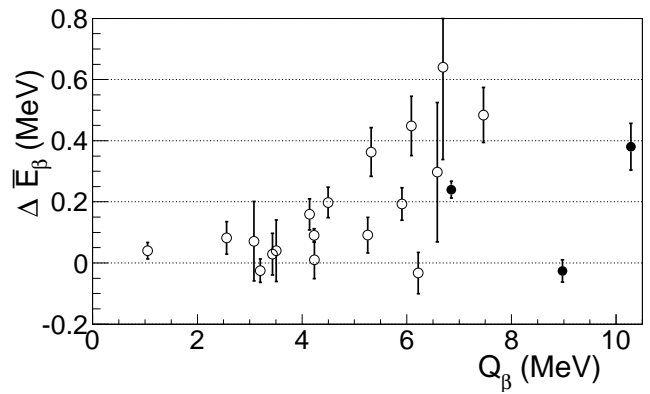}
 \caption{ Difference between average $\beta$ energies obtained by direct $\beta$ spectrum 
measurements (Tengblad {\it et al.} \cite{rud90}) and from TAGS $\beta$ intensity distributions. 
TAGS results are from \cite{gre97} (open circles) and from the present work (filled circles).}
 \label{ebtasteng}
 \end{center}
\end{figure}

Table~\ref{averbet} shows the $\bar{E}_{\beta}$ given in Ref.~\cite{rud90}
obtained from the $\beta$ spectrum measurements of Tengblad {\it et al.}~\cite{ten89}. 
For comparison the average $\beta$ energy obtained from the present
TAGS data, given in Table~\ref{averene},
is incremented with the average $\beta$ energy corresponding
to the $\beta$ delayed neutron branch. 
The contribution of the $\beta$n branch is calculated 
from the $I_{\beta n} (E_{x})$ distribution obtained as
explained in Section~\ref{meas}.
We find
that the values of \cite{rud90} agree with our result for $^{88}$Br but differ by 240~keV for 
$^{87}$Br and by 380~keV for $^{94}$Rb. This situation 
is comparable to that observed
for Greenwood {\it et al.}~\cite{gre97} TAGS data. 
Figure~\ref{ebtasteng} presents in a graphical way the
difference of average $\beta$ energies $\Delta \bar{E}_{\beta}$
between the results of Tengblad {\it et al.} and the results of both
Greenwood {\it et al.} and ourselves. 
In the figure the  differences are represented as a function
of $Q_{\beta}$ to illustrate what seems a systematic trend. 
Although the scattering of values is relatively large,
on average the differences are smaller below $\sim5$~MeV.
The isotopes from Ref.~\cite{gre97} shown in Fig.~\ref{ebtasteng}
are: $^{146}$Ce, $^{145}$Ce, $^{144}$Ba, $^{141}$Ba, $^{143}$La, $^{94}$Sr, 
$^{93}$Sr, $^{145}$La, $^{143}$Ba, $^{89}$Rb, $^{141}$Cs, $^{145}$Ba, $^{91}$Rb, 
$^{95}$Sr, $^{140}$Cs, $^{90}$Rb, $^{90m}$Rb, and $^{93}$Rb, in order of increasing $Q_{\beta}$.

%%%%%  Table 3  %%%%%
\begin{table}[h]
\caption{Comparison of average $\beta$ energies 
obtained from direct $\beta$ spectrum measurement (Tengblad {\it et al.} \cite{rud90})
with those obtained  combining 
$I_{\beta \gamma} (E_{x})$ from present TAGS data and $I_{\beta n} (E_{x})$
derived from neutron spectrum data. See text for details.} 

\begin{center}
%\resizebox{5.7cm}{!}{
\begin{tabular}{ccc} \hline \hline
  & \multicolumn{2}{c}{$\bar{E}_{\beta} \mathrm{(keV)}$} \\ \hline
Isotope& This work &  Ref.~\cite{rud90} \\ \hline
$^{87}$Br &  $1170^{+32}_{-19}$  & $1410 \pm 10$ \\
$^{88}$Br & $1706^{+32}_{-38}$  & $1680 \pm 10$ \\
$^{94}$Rb & $2450^{+32}_{-30}$ & $2830 \pm 70$ \\ \hline \hline
\end{tabular}
%}
\end{center}
\label{averbet}
\end{table}

%%%%%  Figure 6  %%%%%

\begin{figure}[h]
 \begin{center}
 \includegraphics[width=8.6cm]{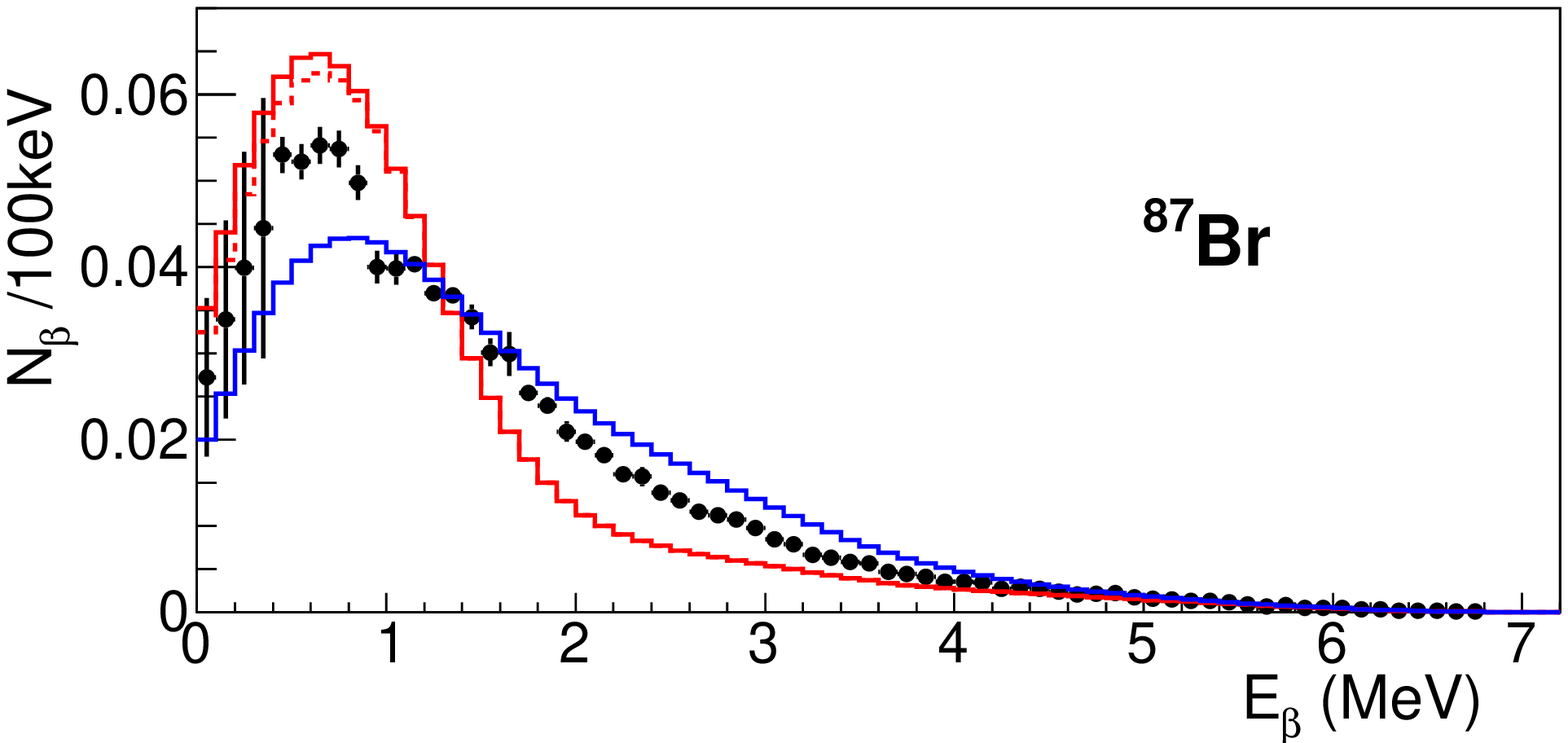}
 \includegraphics[width=8.6cm]{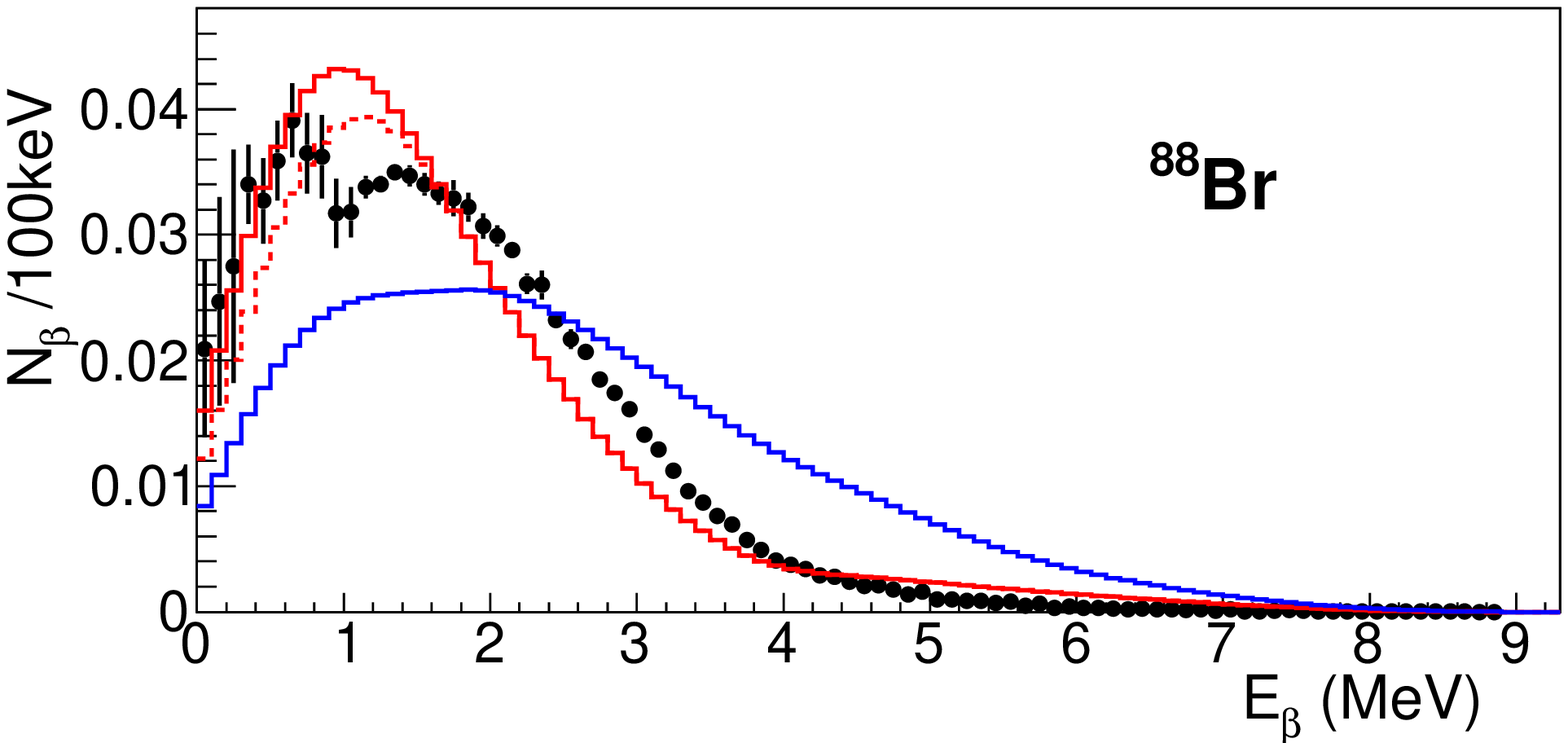}
 \includegraphics[width=8.6cm]{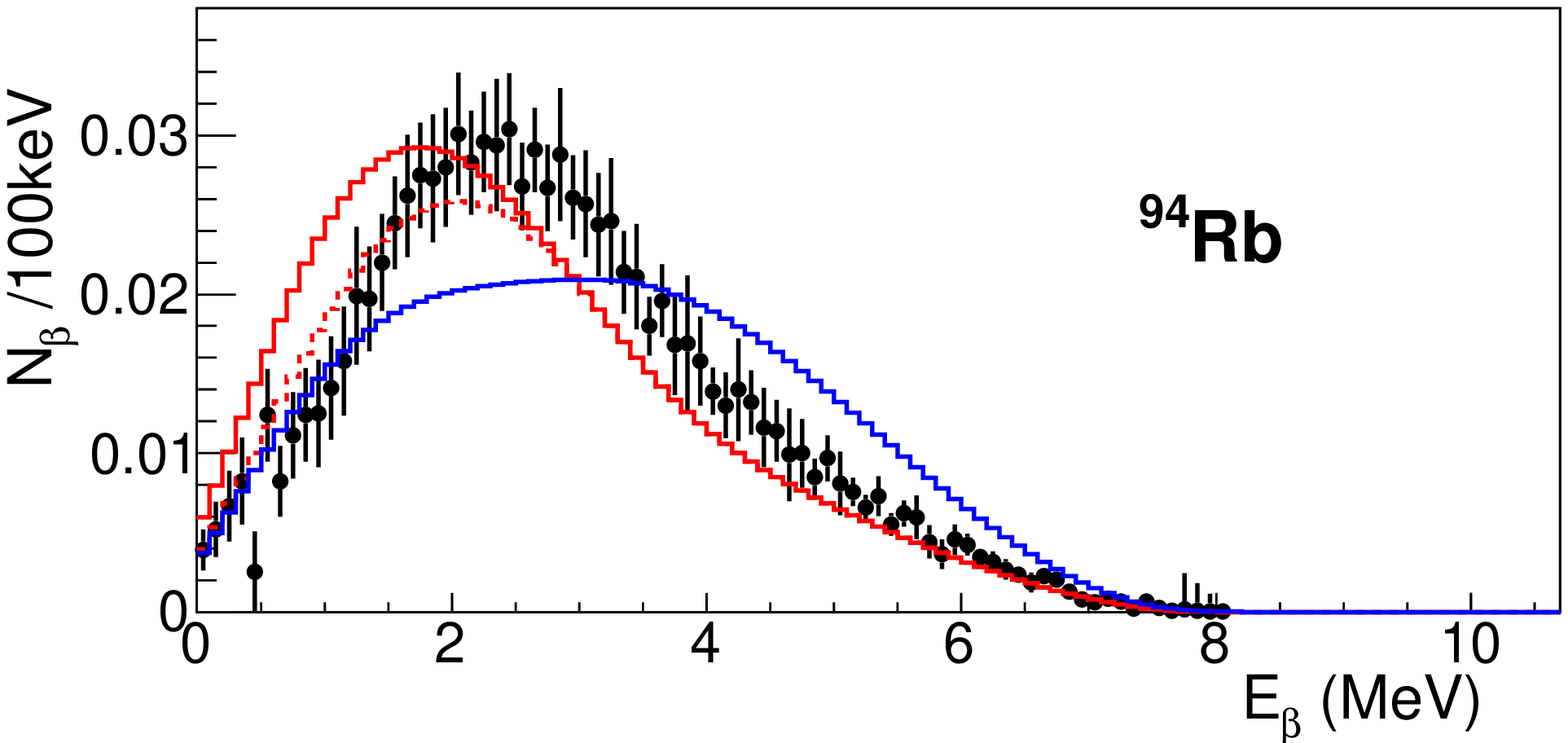}
 \caption{(Color online)  Comparison of $\beta$ spectra $S_{\beta}$. 
Tengblad {\it et al.} \cite{rud90}: black circles; 
present TAGS result: dashed red line; 
present TAGS plus $\beta$ delayed neutron contribution: continuous red line;
high-resolution measurements \cite{br87,br88,rb94}: blue line.}
 \label{betspec}
 \end{center}
\end{figure}

More illustrative than the comparison of average values is the comparison of $\beta$ 
energy distributions $S_{\beta} (E_{\beta})$ as is done in Fig.~\ref{betspec}. 
Large differences in shape between the results of Tengblad {\it et al.}
and the present TAGS results are clearly seen, 
even for $^{88}$Br where the average values agree.
The contribution of the $\beta$-delayed neutron branch, added to the
TAGS result for the comparison, is shown.
For reference we also include in the figure the distribution calculated
from the high-resolution level scheme in ENSDF.
The $S_{\beta} (E_{\beta})$ distribution calculated from the TAGS data is shifted
to lower energies for the three isotopes, in comparison to the direct
$\beta$ spectrum measurement. We should point out that a similar trend
is found for the remaining isotopes included in the same experimental
campaign, $^{86}$Br and $^{91}$Rb~\cite{ric14}, and $^{92,93}$Rb~\cite{zak15a,zak15b},
where we find deviations in $\Delta \bar{E}_{\beta}$ in the range 200 to 400~keV.
Moreover, our results for $^{91}$Rb and $^{93}$Rb agree rather well
with those obtained by Greenwood {\it et al.}~\cite{gre97}.

The assumption of an allowed shape used here to calculate $S_{\beta} (E_{\beta})$
from $I_{\beta} (E_{x})$ introduces some uncertainty in the comparison. 
However it is likely to be a good approximation.
Thus to explain the difference between TAGS results and the direct $\beta$ spectrum measurement
one is forced to consider systematic errors in the use of either one of the two techniques or both.
As explained above we investigated carefully sources of systematic uncertainty
which can lead to distortions of the $\beta$ energy distribution and found
that none of them can explain the observed differences (see Table~\ref{averbet}). 
Moreover as shown in Fig.~\ref{fedmaxerr} the measured TAGS spectrum 
imposes a strong constraint on the bulk of the $\beta$ intensity distribution.
It is difficult to imagine additional sources
of systematic uncertainty which can have a significant impact on the shape of this distribution. 
To clarify the discrepancy new measurements of the spectrum of $\beta$ particles emitted in the decay
of a number of selected isotopes would be of great value.

%%%%%  Figure 7  %%%%%

To finalize this part of the discussion we should point out that $\bar{E}_{\gamma}$
can be obtained from the $\beta$ spectra measured in \cite{ten89}. 
This can be achieved by deconvolution of the $\beta$ spectra
with appropriate $\beta$ shapes $s_{\beta} (Q_{\beta}-E_{x}, E)$
to obtain the $I_{\beta} (E_{x})$ (see Eq.~\ref{eq:betspec}).
As a matter of
fact this procedure is needed (and applied in \cite{ten89}) to obtain the
antineutrino spectrum using Eq.~\ref{eq:nuspec}. The  average $\gamma$ energies obtained
in this way would show systematic differences with respect to TAGS results
of opposite sign to those found for $\bar{E}_{\beta}$. 
Rather than using this approach the authors of
\cite{rud90} determine average $\gamma$ energies $\bar{E}_{\gamma}$
from an independent set of measurements 
using a NaI(Tl) detector to obtain the spectrum of $\gamma$-rays
for the decay of each isotope.
There are also large discrepancies between
these results and those obtained from TAGS measurements. We postpone the
discussion of these differences to a forthcoming publication~\cite{ric14}.

\begin{figure}[h]
 \begin{center}
 \includegraphics[width=8.6cm]{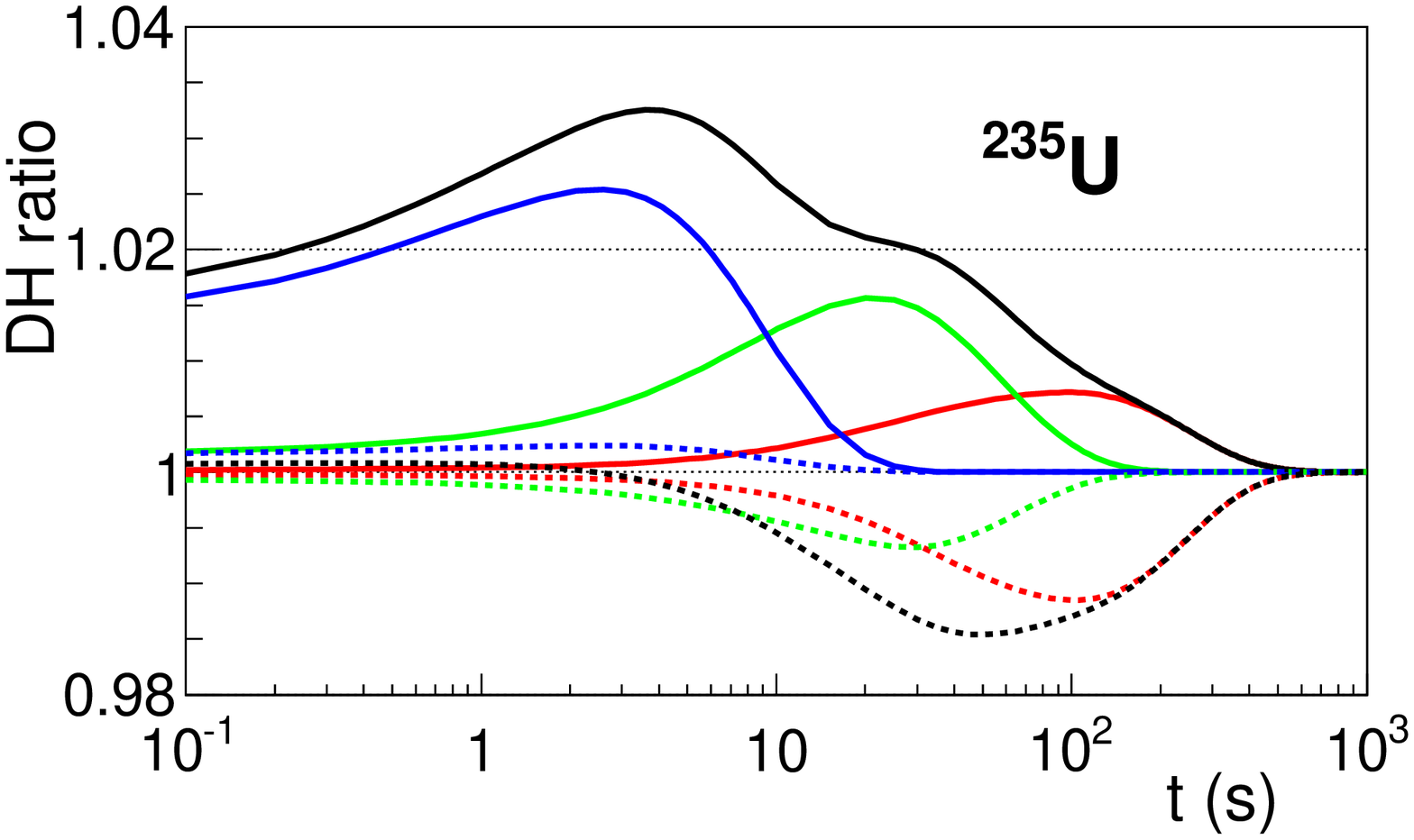}
 \includegraphics[width=8.6cm]{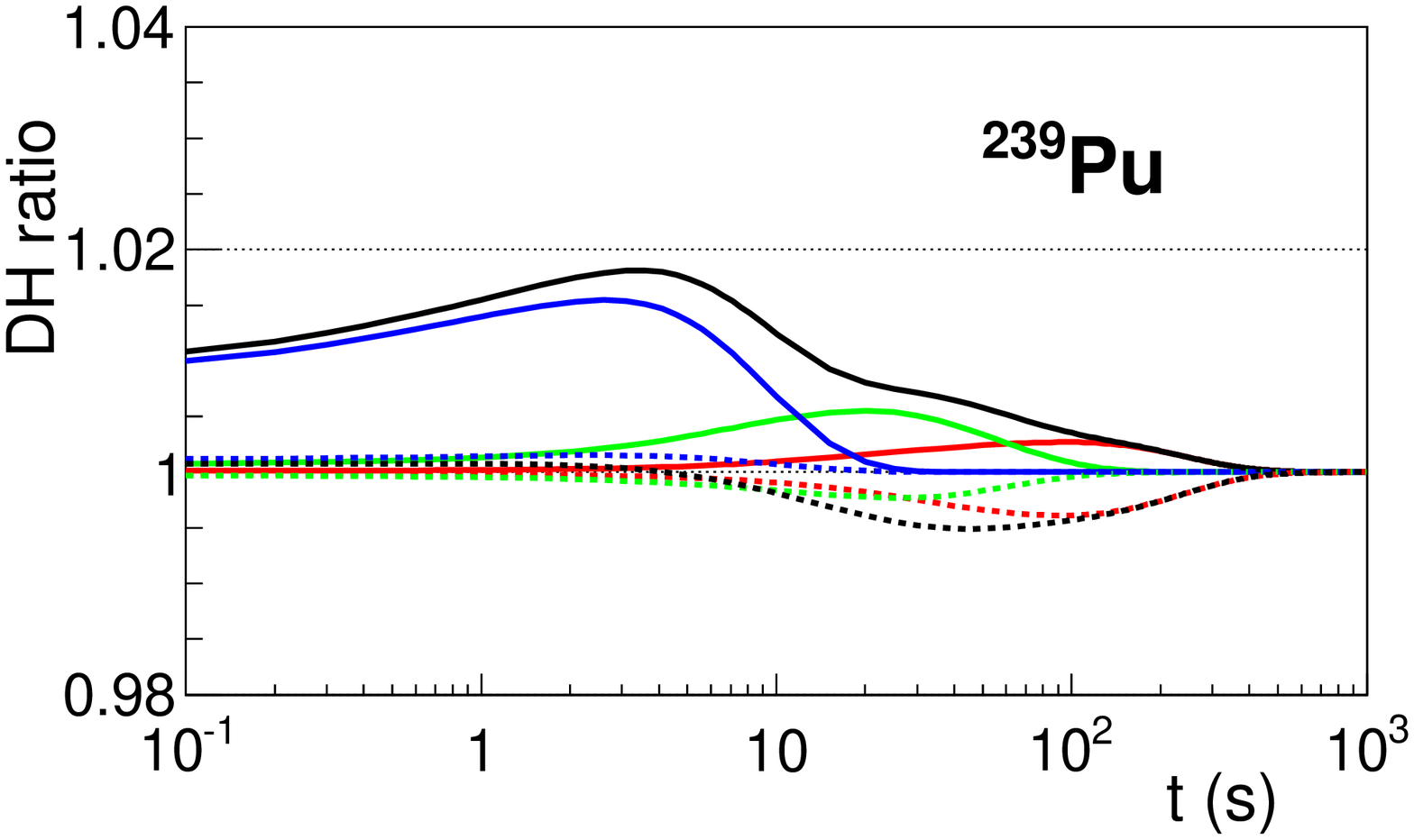}
 \caption{(Color online)  Ratio of decay heat as a function of cooling time
calculated for $^{235}$U and $^{239}$Pu when our TAGS data replaces high-resolution data. 
Continuous line: photon component; dashed line: electron component. 
Red: $^{87}$Br; green: $^{88}$Br; blue: $^{94}$Rb; black: all three isotopes.}
 \label{decayheat}
 \end{center}
\end{figure}

The impact of the present TAGS results for
$\bar{E}_{\gamma}$ and $\bar{E}_{\beta}$ on decay-heat summation calculations was evaluated. 
Figure~\ref{decayheat} shows the ratio of calculations using TAGS
data to calculations using high-resolution data. The figure shows the evolution of the ratio 
as a function of cooling time following the prompt thermal fission
of $^{235}$U and $^{239}$Pu. 
Both together account for most of the power released in most reactors.
The calculation is similar to that described in Ref.~\cite{son15}.
It uses fission yields from JEFF-3.1~\cite{jeff31} and the ENDF/B-VII updated decay data sublibrary.
The update introduces 
$\beta$-intensity distributions from previous TAGS measurements and, for a few isotopes,
from $\beta$-spectrum measurements and from theoretical calculations.
As is customary the DH is evaluated separately for the electromagnetic energy (EEM), 
or photon component
($\gamma$ rays, X rays, \ldots), and for the light particle energy (ELP), or electron component 
($\beta$ particles, conversion electrons, Auger electrons, \ldots).
The ratio is computed for each individual isotope and for the three isotopes together.
As expected the effect of the inclusion of TAGS data is largest for $^{94}$Rb and smallest for $^{87}$Br.
The largest variation in  the EEM component
occurs at short cooling times between 1 and 10~s.
Due to the particular normalization of the high-resolution $^{94}$Rb $\beta$-intensity distribution
mentioned above the effect is not observed in the ELP component (see also Table~\ref{averene}).
The effect is larger for $^{235}$U fission, due to the larger fission yields for the three isotopes, 
reaching an increment of 3.3\% for the combined contribution to the EEM component at $t=3.5$~s.
For $^{239}$Pu the increment reaches 1.8\% . 
Although the impact is somewhat small the present data contribute to
reduce the discrepancy between DH integral measurements and summation calculations
for $^{235}$U in the range of 1 to 100~s (see for example Fig.~12 of Ref.~\cite{jor13}).

%%%%%  Figure 8  %%%%%

\section{Antineutrino spectra
\label{antinu}}

\begin{figure}[h]
 \begin{center}
 \includegraphics[width=8.6cm]{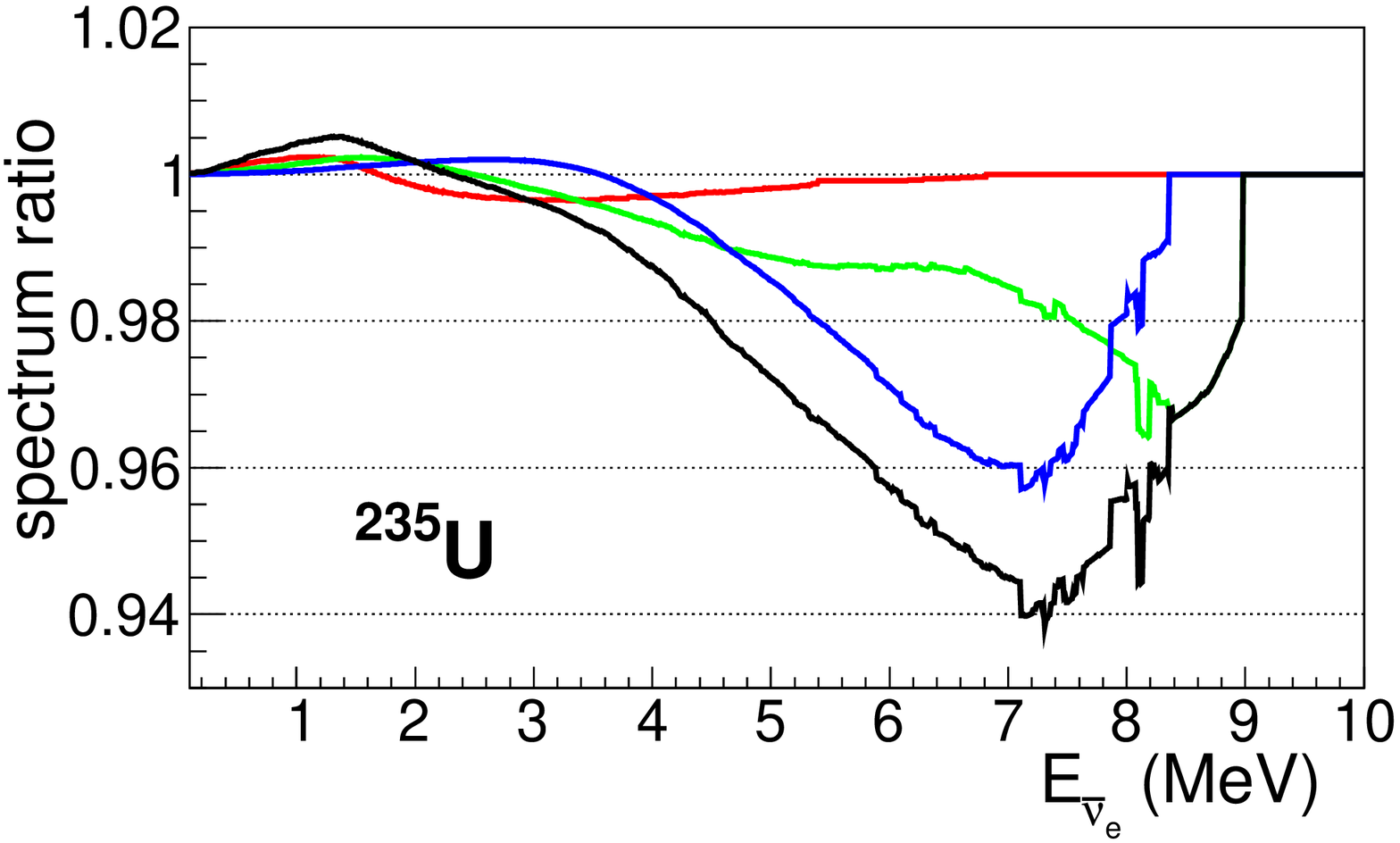}
 \includegraphics[width=8.6cm]{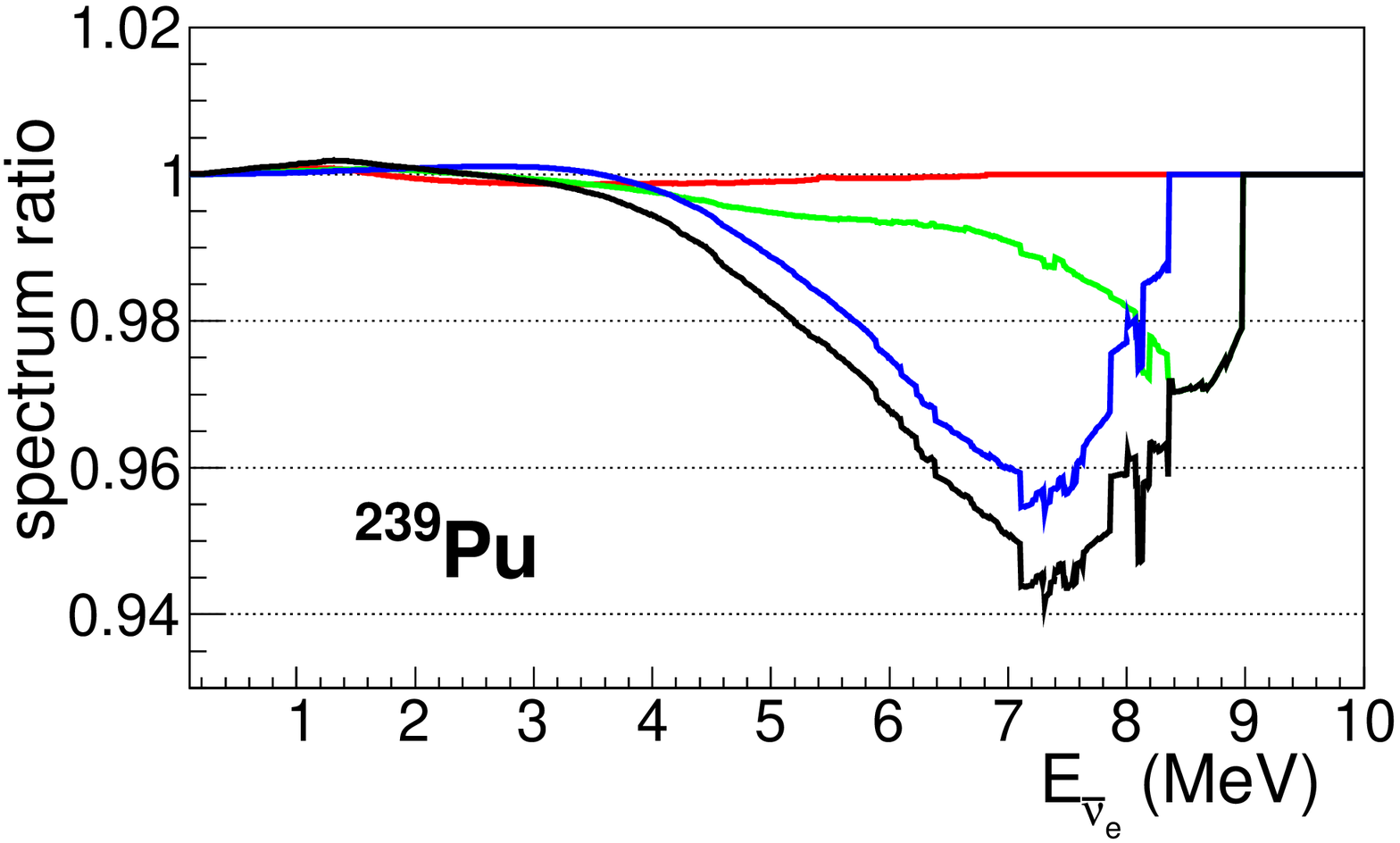}
 \caption{(Color online)  Ratio of antineutrino spectra as a function of energy
calculated for $^{235}$U and $^{239}$Pu when our TAGS data replaces high-resolution data. 
Red: $^{87}$Br; green: $^{88}$Br; blue: $^{94}$Rb; black: all three isotopes.}
 \label{anendfb}
 \end{center}
\end{figure}

\begin{figure}[h]
 \begin{center}
 \includegraphics[width=8.6cm]{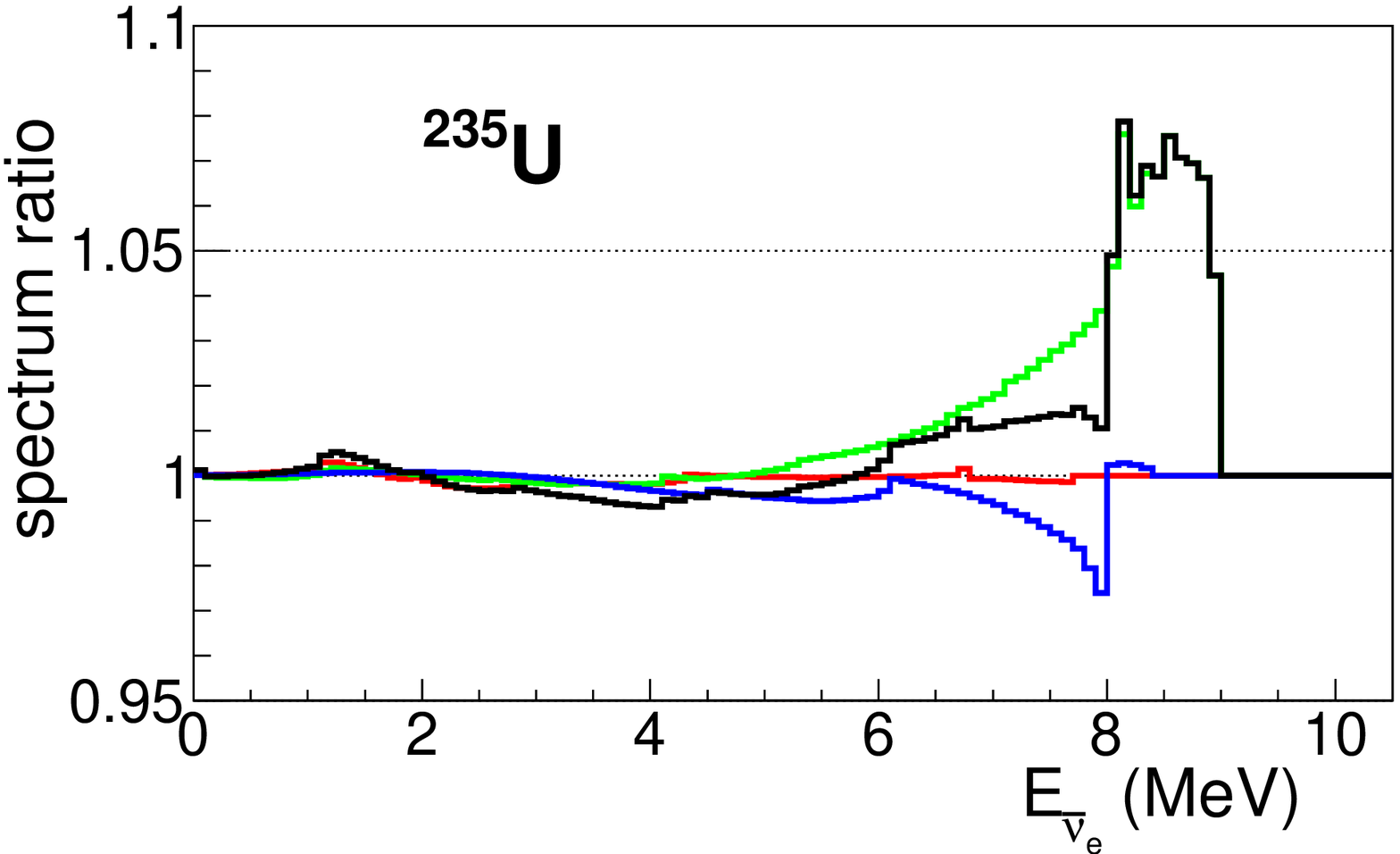}
 \includegraphics[width=8.6cm]{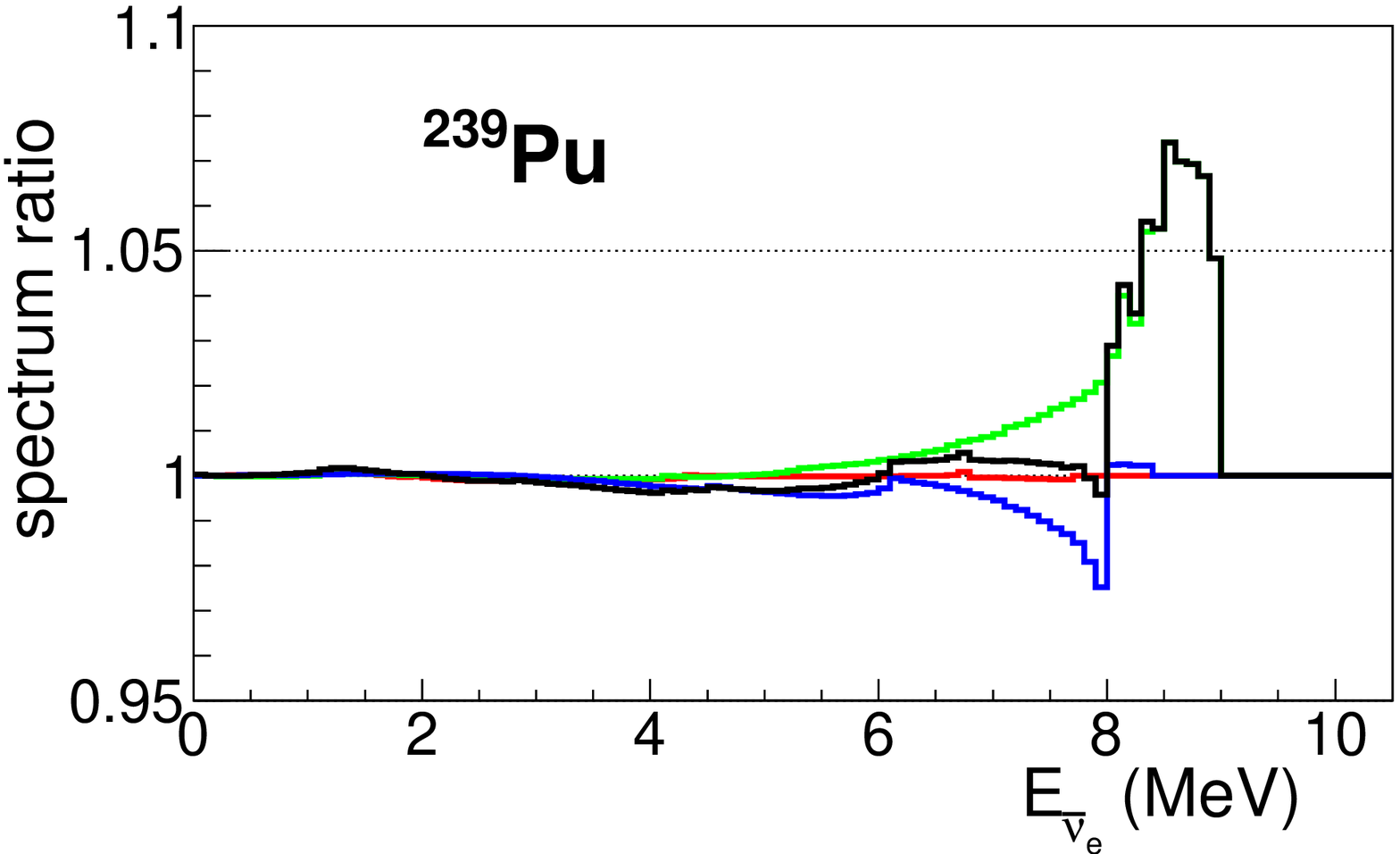}
 \caption{(Color online)  Ratio of antineutrino spectra as a function of energy
calculated for $^{235}$U and $^{239}$Pu when our TAGS data replaces the data 
Tengblad {\it et al.} 
Red: $^{87}$Br; green: $^{88}$Br; blue: $^{94}$Rb; black: all three isotopes.}
 \label{antengblad}
 \end{center}
\end{figure}

The impact of our data on calculated antineutrino spectrum is shown in Fig.~\ref{anendfb}
and Fig.~\ref{antengblad}.
The $\bar{\nu}_{e}$ summation calculation of  Fig.~\ref{anendfb} is analogous to 
the DH calculation of Fig.~\ref{decayheat}.
It shows for $^{235}$U and $^{239}$Pu fission the ratio of calculated $\bar{\nu}_{e}$ spectrum when
our TAGS data replaces the high resolution data.
The effect of each individual isotope and of the three together is shown. 
For both fissioning systems the impact of $^{87}$Br is negligible, 
while the effect of $^{88}$Br peaks around 8.5~MeV (3\%) and 
that of $^{94}$Rb peaks around 7~MeV (4\%). 
The combined effect is a reduction of the calculated $\bar{\nu}_{e}$ spectrum 
which reaches a value of 6\% around 7.2~MeV. Similar figures are 
obtained for $^{238}$U and $^{241}$Pu.
It is remarkable that the effect of our TAGS data for $^{88}$Br 
and $^{94}$Rb is of equal importance to that of
the combined effect of recently measured~\cite{ras16} TAGS data 
for $^{92}$Rb, $^{96}$Y and $^{142}$Cs.
Compare Fig.~\ref{anendfb} in the present work with Fig.~6 of Ref.~\cite{ras16},
which shows an effect of similar shape and magnitude.
These three isotopes contribute most to the  $\bar{\nu}_{e}$ spectrum around
7~MeV, with $^{92}$Rb being the largest contributor~\cite{zak15b}.
Due to current uncertainties in the summation method it is not easy to draw conclusions 
on the impact of both experiments on the origin of the antineutrino spectrum shape distortion.
Note that they lead to a {\it reduction} of the calculated spectrum which is maximum 
about 1~MeV above the center of the observed {\it excess}. 
Better quality data for a larger set of isotopes, including decay data and fission yields, is required.
Our result shows the importance of performing TAGS measurements for fission products 
with very large $Q_{\beta}$-value, which are likely to be affected by large $Pandemonium$
systematic error, even if they have moderate fission yields.

%%%%%  Figure 9  %%%%%

Figure \ref{antengblad} shows a different set of $\bar{\nu}_{e}$ summation calculations.
The calculation is analogous to that described in Ref.~\cite{fal12}. 
It uses a different selection of decay data from the calculation shown in Fig.~\ref{anendfb}.
More specifically it uses antineutrino spectra derived from the $\beta$ 
spectra of Tengblad {\it et al.}~\cite{ten89} 
for $^{87,88}$Br and $^{94}$Rb instead of $\bar{\nu}_{e}$ spectra derived from high-resolution data.
Thus Fig.~\ref{antengblad} shows the effect of replacing Tengblad {\it et al.} data with our TAGS data.
As can be seen the replacement of $^{87}$Br has little impact, while there is a cancellation
below $E_{\bar{\nu}_{e}}=8$~MeV between the $^{88}$Br and $^{94}$Rb deviations. 
However the difference
between our TAGS data and the data of Tengblad {\it et al.} for $^{88}$Br produces an increase
in the calculated antineutrino spectra of about 7\% between 8 and 9~MeV.
Note that although $^{94}$Rb has a $Q_{\beta}$ of 10.28~MeV we do not observe appreciable
$\beta$ intensity below 2.41~MeV excitation energy, thus the maximum 
effective endpoint energy is below 8~MeV.
The relatively large impact of $^{88}$Br is due to the fact that only a few decay 
branches contribute to the spectrum here.
Note that in this energy interval the uncertainty of the integral $\beta$-spectrum 
measurements~\cite{sch85,hah89} is relatively large, thus summation 
calculations are particularly relevant.
This points again to the need to perform TAGS measurements for fission
products with very large $Q_{\beta}$.

%%%%%  Figure 10  %%%%%

\section{Gamma intensity from neutron unbound states
\label{gamneu}}

Figure~\ref{tasint} shows for all three isotopes 
a sizable TAGS intensity $I_{\beta \gamma} (E_{x})$ above $S_{n}$.
This intensity extends well beyond the first few hundred keV
where the low neutron penetrability makes $\gamma$-ray emission competitive.
For comparison the figure includes 
the $\beta$-intensity distribution followed by neutron emission 
$I_{\beta n} (E_{x})$ deduced from the neutron spectrum as explained above. 
The integrated decay intensity above $S_{n}$ followed
by $\gamma$-ray emission $P_{\gamma} = \int_{S_{n}}^{Q_{\beta}} I_{\beta \gamma} (E_{x}) dE_{x}$  
obtained from the TAGS measurement
is compared to the integrated $I_{\beta n} (E_{x})$
or $P_{n}$ value in Table~\ref{gamneu}. 
Surprisingly large values of $P_{\gamma}$ are obtained, which in the case of $^{87}$Br is
even larger than $P_{n}$. The $\gamma$ branching represents
57\% of the total for $^{87}$Br, 20\% for $^{87}$Br and 4.5\% for $^{94}$Rb.
In the case of $^{87}$Br we find 8 times more intensity 
than the high-resolution  experiment~\cite{ram83},  which can be explained
by the {\it Pandemonium} effect. The quoted uncertainty on the 
TAGS integrated intensity $P_{\gamma}$ is completely dominated by systematic uncertainties
since the uncertainty due to data statistics is below 0.6\% (relative value) in
all cases. 

%%%%%  Table 4  %%%%%
\begin{table}[h]
\caption{Integrated $\beta$-intensity $P_{\gamma}$ from TAGS data above $S_n$ 
compared to $P_{n}$ values from \cite{br87,br88,rb94}.}

\begin{center}
%\resizebox{5.7cm}{!}{
\begin{tabular}{ccc} \hline
Isotope &  $ P_{\gamma}$ & $P_{n}$  \\
	& (\%) & (\%)  \\ \hline
$^{87}$Br & $3.50^{+49}_{-40}$ & 2.60(4)  \\
$^{88}$Br & $1.59^{+27}_{-22}$ &  6.4(6)  \\
$^{94}$Rb & $0.53^{+33}_{-22}$ & 10.18(24)  \\ \hline
\end{tabular}
%}
\end{center}
\label{gamneu}
\end{table}

We have evaluated several sources of systematic uncertainty.
In the first place we consider uncertainties that affect the overall $\beta$ intensity distributions, 
which were already detailed in Section~\ref{anal}.
To quantify the uncertainties in $P_{\gamma}$ coming from the spread of possible solutions
compatible with the data (see Fig~\ref{fedmaxerr}) we follow the approach
used in Section~\ref{heat} and take the maximum positive and negative difference
with respect to the adopted solution as a measure of this uncertainty.

In addition to this we consider other sources of uncertainty which mostly affect the integral value.

A possible source of uncertainty is related to
the correlations introduced by the finite energy resolution
in the deconvolution process. This can cause a relocation of counts 
in a region of rapidly changing intensity~\cite{tai07b}, such as the region around $S_{n}$. 
However we estimate
from a model deconvolution that this effect is not relevant in the present case.
Likewise the uncertainty on width calibration also has an impact on the redistribution of counts
around $S_{n}$. The highest width calibration point is at 4.123~MeV. From the comparison
of different fits, varying the number and distribution of calibration points, we determine
that the extrapolation of the calibration curve can vary by up to $\pm 15$\% at 10~MeV.
This introduces an uncertainty in $P_{\gamma}$ of 2\% for $^{87}$Br and 6\% for
$^{88}$Br and $^{94}$Rb.

The uncertainty in the energy calibration of TAGS spectra might have
an impact on the result because 
of the dependence of the response on energy. 
However we verified that this effect is negligible. 
The main effect of the uncertainty on the energy calibration is on
the integration range. Since the intensity is rapidly changing in the region around $S_{n}$
the effect can be large. The fact that the structure observed
in the distribution of Fig.~\ref{gamtot} around $7-8$~MeV for $^{94}$Rb
coincides with the levels populated
in the final nucleus (see next Section) allows us to conclude that  the
energy calibration at $S_{n}$ is correct to about one energy bin (40~keV).
We evaluate the uncertainty in the integral,
equivalent to changes of half a bin, to be 11\% for the bromine isotopes and 15\%
for rubidium. 

The uncertainty values entered in Table~\ref{gamneu} correspond to the sum in quadrature 
of the three types of uncertainty mentioned above: uncertainties in the deconvolution, and uncertainties
in the resolution and energy calibration.

\section{Comparison with Hauser-Feshbach calculations 
\label{calc}}

\begin{figure}[h]
 \begin{center}
 \includegraphics[width=8.6cm]{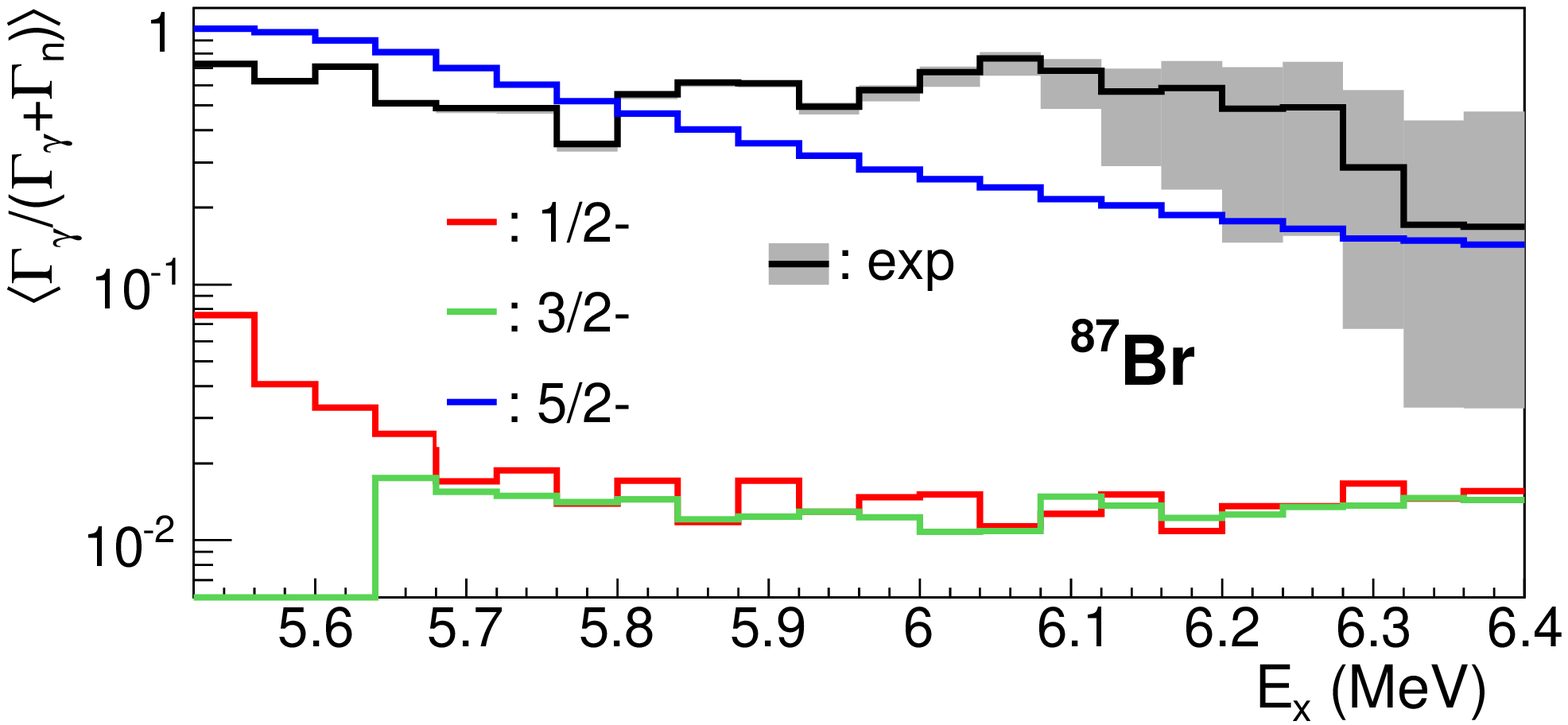}
 \includegraphics[width=8.6cm]{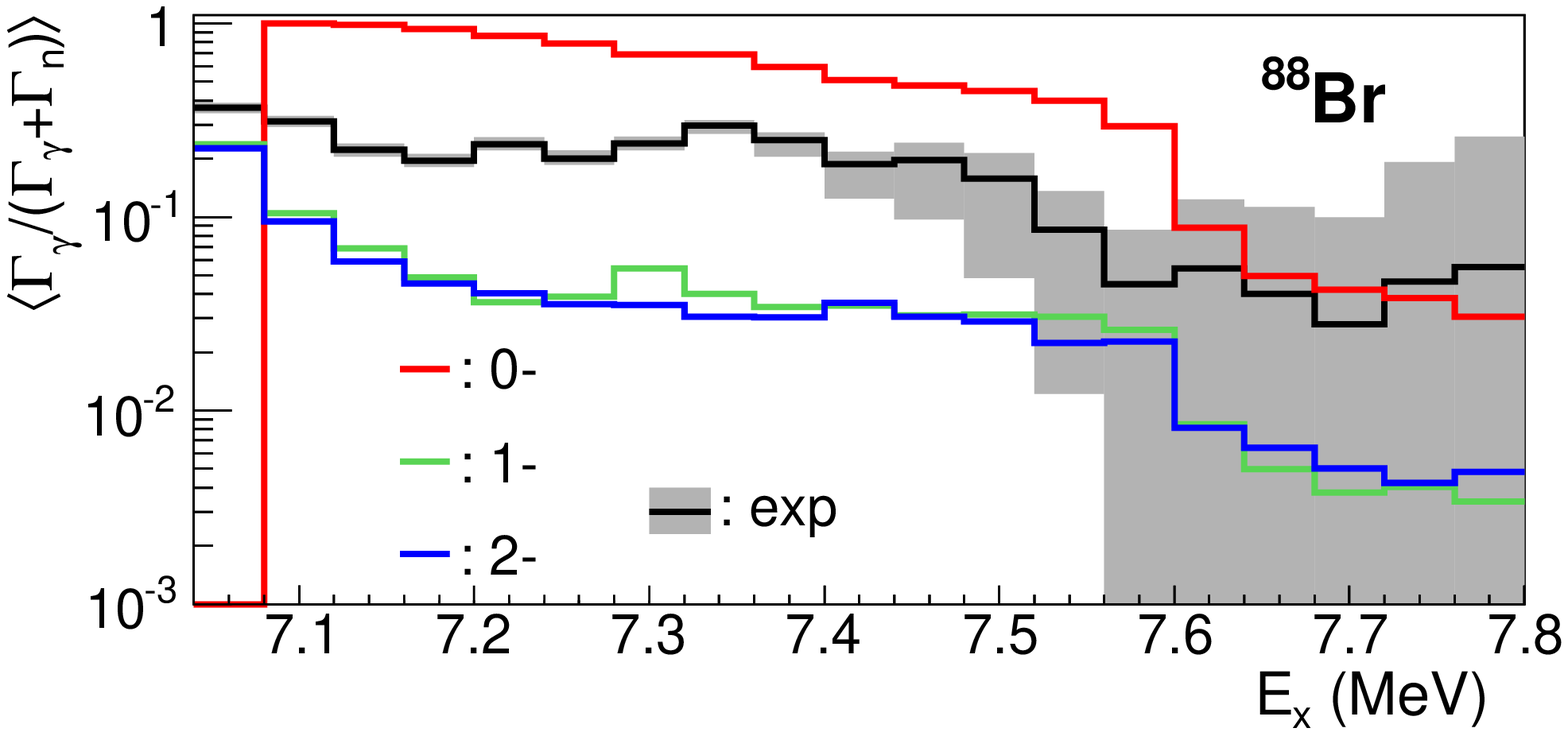}
 \includegraphics[width=8.6cm]{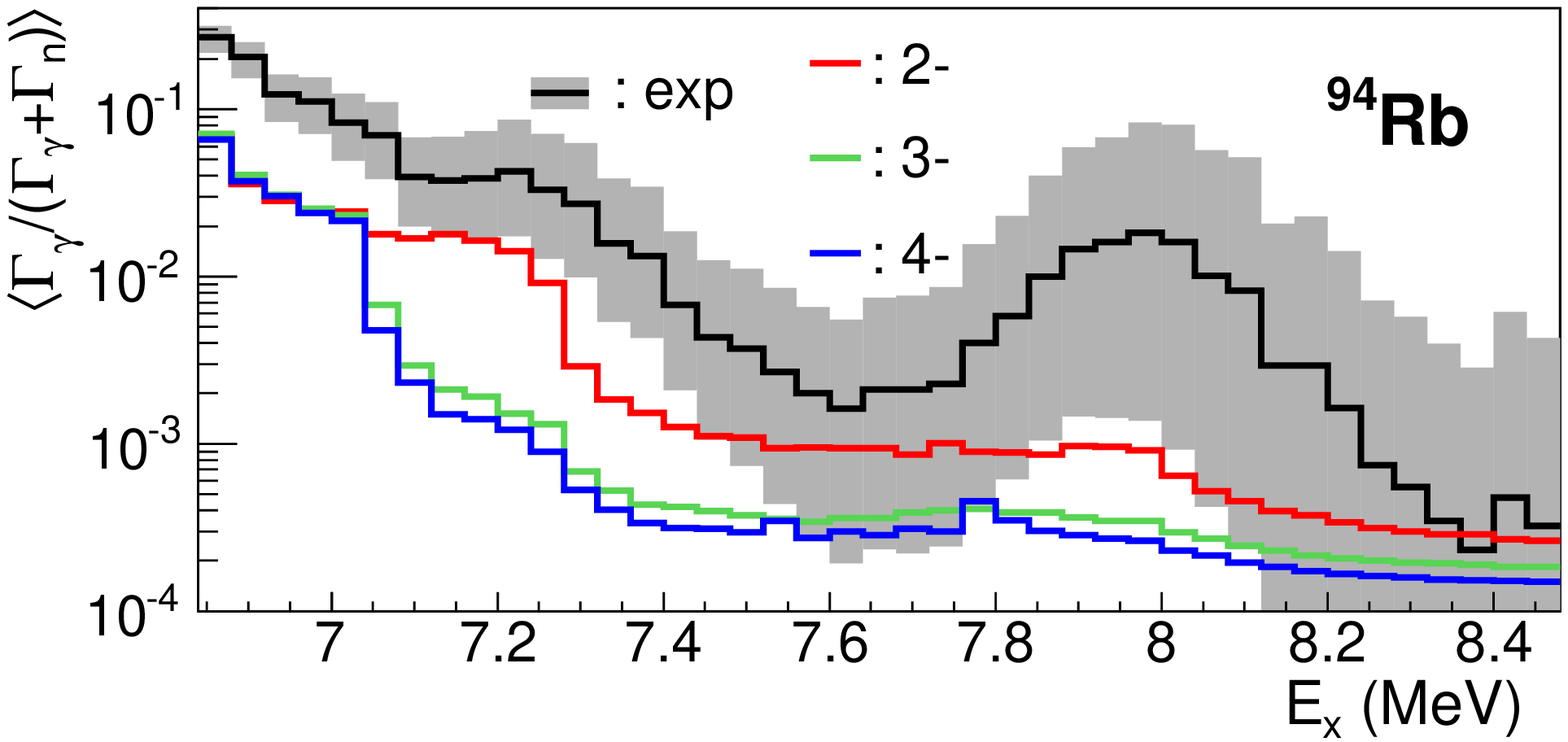}
 \caption{(Color online)  Average gamma to total width from experiment (black line) and calculated for
the three spin-parity groups populated in allowed decay (red, green, blue). The 
gray-shaded area around the experiment indicates the sensitivity to systematic effects.
See text for details.}
 \label{gamtot}
 \end{center}
\end{figure}

We show in Fig.~\ref{gamtot} the ratio 
$I_{\beta \gamma} (E_{x})/ (I_{\beta \gamma} (E_{x}) + I_{\beta  n} (E_{x}))$
as a function of excitation energy.
The shaded area represents the uncertainty in the ratio
coming from the spread of solutions $I_{\beta \gamma} (E_{x})$ to the TAGS inverse problem
shown in Fig.~\ref{fedmaxerr}.
It should be noted that the ratio is affected also
by systematic uncertainties in the $I_{\beta n} (E_{x})$ distribution coming
from the deconvolution of neutron experimental spectra
as well as by uncertainties in the neutron spectra themselves,
but they are not considered here.

The experimental intensity ratio 
in Fig.~\ref{gamtot} is identical to the 
average ratio $\langle \Gamma_{\gamma}(E_{x})/\Gamma_{tot}(E_{x}) \rangle $.
The average is taken over all levels in each bin populated in the decay. 
Thus the experimental distribution can be directly compared  with the results
of Hauser-Feshbach calculations of this ratio. 
The NLD and PSF used in the calculations
are the same as used to construct  
the spectrometer response to the decay (see Section~\ref{anal}).
The new ingredient needed is the NTC which is obtained from the Optical Model (OM).
It is calculated with Raynal's ECIS06 OM code integrated in the TALYS-1.4 software
package~\cite{talys}. OM parameters are taken from the so-called local
parametrization of Ref.~\cite{kon03}.
Neutron transmission is calculated for final levels known to be populated in the decay:
g.s. of $^{86}$Kr, g.s and first excited state of $^{87}$Kr, and g.s. plus 8 excited states of
$^{93}$Sr. 
With these ingredients one obtains the average widths 
$\langle \Gamma_{\gamma} \rangle$ and $\langle \Gamma_{n} \rangle$
(see Appendix).

In the case of $^{87}$Kr we can 
compare the calculated average values
with experimental data obtained from neutron 
capture and transmission reactions~\cite{ram83,car88}.
In particular for $1/2^{-}$ and $3/2^{-}$ resonances which are
populated in the decay of a $3/2^{-}$ $^{87}$Br ground state.  
Up to fifty $1/2^{-}$ and sixty-six $3/2^{-}$ resonances were identified in an interval of
960~keV above $S_{n}$. The NLD of Ref.~\cite{gor08} predicts 46 and 90 respectively, in
fair agreement with these values. 
The distribution of neutron widths for $1/2^{-}$ resonances 
in the interval $E_{n} = 250-960$~keV
is compatible with a PT distribution with average width
$\langle  \Gamma_{n} \rangle = 1.95$~keV. 
The same is true for $3/2^{-}$ resonances with $\langle  \Gamma_{n} \rangle = 2.79$~keV.
In the same interval the Hauser-Feshbach calculated widths vary 
between 0.3~keV and 0.7~keV for  $1/2^{-}$ states and
between 0.5~keV and 0.9~keV for $3/2^{-}$ states. 
In both cases the calculation is about 4 times too low. 
The information on $\langle \Gamma_{\gamma} \rangle$ is less abundant. The
$\gamma$ width has been determined for six $1/2^{-}$ and ten $3/2^{-}$ resonances,
with values in the range 0.075-0.48~eV, and is fixed to 0.255~eV, from
systematics, for the remaining resonances. The Hauser-Feshbach calculation
gives values in the range 0.08-0.12~eV. On average the calculation is about a factor
three too low. Since the NLD reproduces the number of resonances, to reach
such values for the partial widths requires a renormalization by a factor of 3-4 for
the PSF and the NTC in $^{87}$Kr, which seems large.
The reader should note that variations of similar magnitude and direction
for both the PSF and NTC have little impact on the calculated ratio
$\langle \Gamma_{\gamma}/\Gamma_{tot} \rangle$.
It should also be noted that this ratio
is insensitive to changes in NLD.

%%%%%  Figure 11  %%%%%

We show in Fig.~\ref{gamtot} the ratio  $\langle \Gamma_{\gamma}/\Gamma_{tot} \rangle$,
calculated with nuclear statistical parameters as described above,
for the three spin-parity
groups populated under the Gamow-Teller selection rule.
Due to statistical fluctuations affecting individual widths~\cite{jon76}, 
this cannot be obtained as 
$\langle \Gamma_{\gamma} \rangle / (\langle \Gamma_{\gamma}\rangle  + \langle \Gamma_{n} \rangle )$.
Rather than trying to obtain a formula for the average correction factor to be applied
to this ratio, 
which is the common practice for cross section calculations~\cite{talys},
we use the
Monte Carlo method to obtain directly the average of width ratios.
The procedure to obtain a statistical realization (or sample) from the model
is similar to that described in Ref.~\cite{tai07a}.
Level energies for each spin-parity are generated
according to a Wigner distribution from the NLD. 
For each state the corresponding 
$\Gamma_{\gamma}$ and $\Gamma_{n}$ to individual final states
are sampled from PT distributions
with the calculated average values (see Appendix).
The total $\gamma$ and neutron widths are obtained 
by summation over all possible final states and the ratio is
computed. The ratio is averaged for all levels lying
within each energy bin (40~keV). In order to eliminate fluctuations  
in the calculated averages, the procedure is repeated
between 5 and 1000 times depending on level density.
%It should be noted that in the case of $^{87}$Br 
%where the level density is quite low the actual value
%can fluctuate significantly around the calculated average.
Very large average enhancement factors are obtained, reaching
two orders-of-magnitude, when the neutron emission is dominated
by the transition to a single final state. 

In the case of the decay of the $3/2^{-}$ ground state in $^{87}$Br   
one can see in Fig.~\ref{gamtot} that the
strong $\gamma$-ray emission above $S_{n}$ 
can be explained as a consequence of the large hindrance
of $l=3$ neutron emission from $5/2^{-}$ states in $^{87}$Kr 
to the $0^{+}$ g.s. of  $^{86}$Kr. 
This is the explanation already proposed in \cite{ram83}.
The situation is even more favorable to this explanation
if the spin-parity of $^{87}$Br were $5/2^{-}$ as suggested
in \cite{por06}. In this case the neutron emission is hindered for both 
$5/2^{-}$ and $7/2^{-}$ states populated in the allowed decay.
In the case of $^{88}$Br $1^{-}$ decay a
similar situation occurs for $0^{-}$ states in $^{88}$Kr
below the first excited state in $^{87}$Kr at 532~keV,
which requires  $l=3$ neutron emission to populate the $5/2^{+}$
g.s. in $^{87}$Kr. 
It should be noted that if 
the spin-parity of $^{88}$Br were $2^{-}$ as suggested in~\cite{br88} 
the three allowed spin-parity groups $(1^{-},2^{-},3^{-})$ will have similar
gamma-to-total ratios, a factor of 3 to 5 too low compared
to experiment, which reinforces our choice of $1^{-}$ for the $^{88}$Br g.s. 
A more quantitative comparison of the experimental and
calculated ratios requires a knowledge of the distribution of $\beta$ intensity
between the three spin groups.
This can be obtained from $\beta$ strength theoretical
calculations, such as those in \cite{moe03} for example.
It is clear however that for both bromine isotopes the large
$\gamma$ branching above $S_{n}$ can be explained
as a nuclear structure effect: the absence of states in
the final nucleus which can be populated 
through the emission of neutrons of
low orbital angular momentum.

The case of $^{94}$Rb $3^{-}$ decay is
the most interesting. 
The final nucleus $^{93}$Sr 
is five neutrons away from $\beta$ stability.
Although the $\gamma$ intensity is strongly reduced,
only 5~\% of the neutron intensity, is detectable 
up to more than 1~MeV above $S_{n}$. The structure observed
in the distribution of the average ratio $\langle \Gamma_{\gamma}/\Gamma_{tot} \rangle$,
can be associated
with the opening of $\beta$n channels to different
excited states in $^{93}$Sr.
As can be seen the structure is reproduced by the calculation, 
which confirms the energy calibration at high excitation energies.
In any case the
calculated average gamma-to-total ratio is well
below the experimental value. 
In order to bring the calculation in line with
the experimental value one would need to enhance
the $\gamma$ width, or suppress the neutron width, or any suitable combination of the two, 
by a very large factor of about one order-of-magnitude. 
A large enhancement of the $\gamma$ width,  
and thus of the calculated $(\mathrm{n},\gamma)$ cross sections,
would have an impact on $r$ process abundance calculations~\cite{gor98,sur01,arc11}.
It would be necessary to confirm the large enhancement of the 
$\langle \Gamma_{\gamma}/\Gamma_{tot} \rangle$ ratio observed
in $^{94}$Rb with similar studies on other neutron-rich nuclei in this mass region
as well as in other mass regions. It will also be important to quantify
the contribution of a possible suppression of the neutron width to the observed ratio.

\section{Summary and Conclusion
\label{conclusion}}

We apply the TAGS technique to study the decay of three $\beta$-delayed
neutron emitters. For this we use a new segmented $\mathrm{BaF}_{2}$ spectrometer
with reduced neutron sensitivity, which proved to be well suited to this purpose.
The three isotopes, $^{87}$Br, $^{88}$Br and $^{94}$Rb, 
are fission products with impact in reactor decay heat and
antineutrino spectrum summation calculations.
We obtain $\beta$ intensity distributions which are free from the
$Pandemonium$ systematic error, affecting the 
data available in the ENSDF data base for the three isotopes.
The average $\gamma$-ray energies that we obtain are 31\%, 59\% and 235\%
larger than those calculated with this data base for $^{87}$Br, $^{88}$Br and $^{94}$Rb
respectively, while the average $\beta$ energies are 28\%, 33\% and 13\% smaller.

We compare the energy  distribution of $\beta$ particles emitted in the decay
derived from our $\beta$ intensity distributions with the direct $\beta$ spectrum
measurement performed by Tengblad {\it et al.}, and find significant discrepancies.
Our distributions are shifted to somewhat lower energies. 
This is reflected in the average $\beta$ energies, which we find to be
17\% and 13\% smaller for $^{87}$Br and $^{94}$Rb respectively.
Similar systematic differences are found when the TAGS data of Greenwood {\it et al.} for 18 isotopes
is compared with the data of Tengblad {\it et al.}.
We performed a thorough investigation of possible systematic errors
in the TAGS technique and find that none of them can explain
the observed differences.
It will be important to perform new direct measurements
of the $\beta$ spectrum for a few selected isotopes in
order to investigate this issue further.

We estimate the effect of the present data on DH summation
calculations. We find a relatively modest impact when the high resolution decay data are replaced
by our TAGS data. The impact in the photon component is largest at short cooling
times. For $^{235}$U thermal fission it reaches an increment of 3.3\% around 3.5~s after
fission termination.  This is mainly due to the decay of $^{94}$Rb. 
The influence of $^{88}$Br is smaller and peaks at around 25~s.
In spite of being small it contributes to reduce the discrepancy between
DH integral measurements of the EEM component and summation 
calculations for $^{235}$U in the range
of 1 to 100~s. Many FP contribute in this time range, thus additional TAGS measurements of short
lived FP are required to remove the discrepancy.
In the case of $^{239}$Pu the maximum increment is about 1.8\%.

We also evaluate the impact of the new TAGS data on antineutrino spectrum summation calculations.
When our data replace the data from high-resolution measurements we observe a reduction
of the calculated $\bar{\nu}_{e}$ spectrum which reaches a maximum value of 6\% at 7~MeV
for the thermal fission $^{235}$U. A similar value is obtained for $^{239}$Pu. The reduction is mainly
due to the decay of $^{94}$Rb. The effect of $^{88}$Br, somewhat smaller, peaks at 8.5~MeV.
It is remarkable that we find an impact similar to that observed for $^{92}$Rb, $^{96}$Y  
and $^{142}$Cs which make the largest contribution to the antineutrino spectrum at these energies.
The reason is that the large value of the {\it Pandemonium} systematic error prevails
over the relatively small fission yield for the isotopes studied in this work.
We also verified the effect of replacing our TAGS data with Tengblad {\it et al.} $\beta$-spectrum data.
We found a relatively small impact below $E_{\bar{\nu}_{e}} = 8$~MeV in part 
due to a compensation effect
of the deviations for $^{94}$Rb and $^{88}$Br. However between 8 and 9~MeV
the use of TAGS data for $^{88}$Br leads to an increase of about 7\% in 
the calculated antineutrino spectrum.
This relatively large impact is due to the small number of decay branches in this energy range.
All this underlines the need for TAGS measurements for fission products with a very
large $Q_{\beta}$ decay energy window.

We confirm the suitability of the TAGS technique for obtaining
accurate information on $\gamma$-ray emission from neutron-unbound states.
In order to assess the reliability of the result we examined the systematic errors
carefully since they dominate the total uncertainty budget.
Surprisingly large $\gamma$-ray branchings of  57\% and 20\%  
were observed for  $^{87}$Br and  $^{88}$Br respectively. In the case of $^{94}$Rb
the measured branching of 4.5\% is smaller but still significant.
For $^{87}$Br
we observe 8 times more intensity than previously detected with high resolution $\gamma$-ray
spectroscopy, which confirms the need to use the TAGS technique for such studies.

Combining the information obtained from TAGS measurements
about the $\gamma$ intensity from states  above $S_{n}$ 
with the $\beta$-delayed neutron intensity we can
determine the branching ratio 
$\langle \Gamma_{\gamma}/(\Gamma_{\gamma} + \Gamma_{n}) \rangle$ as a function of $E_{x}$.
The information 
thus acquired, can be used to constrain the neutron capture cross-section
for unstable neutron-rich nuclei.
This opens a new field for applications of $\beta$-decay TAGS studies. 
It also provides additional arguments for the need for accurate measurements 
of $\beta$-delayed neutron emission 
in exotic nuclei. The measurements should cover 
neutron spectra and yields as  well as neutron-$\gamma$ coincidences.

From the comparison of our experimental results with Hauser-Feshbach calculations
we conclude that the large $\gamma$ branching observed in $^{87}$Br and $^{88}$Br
is a consequence of the nuclear structure. Some of the resonances
populated in the decay can only disintegrate via the emission of a 
kinematically hindered neutron to the levels available in the final nucleus.
A similar situation can occur for other $\beta$-delayed neutron emitters,
when the number of levels in the final nucleus 
within the emission window $Q_{\beta} - S_{n}$ is small.
It should be noted that such strong $\gamma$ to neutron competition
introduces a large correction to the estimation of $\beta$-delayed neutron
emission probabilities from $\beta$-strength calculations and should
be taken into account when comparing experiment with calculation.

The case of $^{94}$Rb, is more representative
of the situation expected for nuclei far from stability,
where many levels are available thus the decay by low $l$ neutron emission is always possible. 
For $^{94}$Rb we find that the $\gamma$-ray emission from neutron-unbound states
is largely suppressed, but still 
much larger (an order-of-magnitude) than the result of Hauser-Feshbach
calculations using standard parameters for level density, photon strength and neutron transmission.
If such enhancement with respect to the Hauser-Feshbach model
is due mainly to an increment in the radiative width,  then a similar increase
is obtained for the neutron capture cross-section.
This can have a significant impact on calculated elemental abundances in the astrophysical $r$ process.
It is necessary to confirm and generalize the result obtained
for the neutron-rich nucleus $^{94}$Rb extending this type of study to other 
$\beta$-delayed neutron emitters in the same and different mass regions,
in particular farther away from the valley of $\beta$-stability.
Such measurements using the TAGS technique are already underway and 
additional studies are planned.

\begin{acknowledgments}
This work was supported by Spanish Ministerio de Econom\'{\i}a y
Competitividad under grants FPA2008-06419, FPA2010-17142, FPA2011-
24553, FPA2014-52823-C2-1-P, CPAN CSD-2007-00042 (Ingenio2010)
and the program Severo Ochoa (SEV-2014-0398). 
WG would like to thank the University of Valencia for support.
This work was supported by the Academy of Finland under the Finnish Centre of
Excellence Programme 2012-2017 (Project No. 213503, Nuclear and Accelerator-Based
Physics Research at JYFL).
Work supported by EPSRC(UK) and STFC(UK).
Work partially supported by the European Commission under
the FP7/EURATOM contract 605203.
We thank D. Lhuillier for making available in digital form data tabulated in Ref.~\cite{rud90}.
\end{acknowledgments}

\appendix
\section*{Appendix
\label{appendix}}

The average $\gamma$ width for initial levels (resonances) of spin-parity $J^{\pi}_{i}$
at excitation energy $E_{x}$ can be obtained by summation over all final states
of spin-parity $J^{\pi}_{f}$ and excitation energy $E_{x}-E_{\gamma}$:

%\begin{widetext}
\begin{equation}
\langle \Gamma_{\gamma} (J^{\pi}_{i},E_{x}) \rangle = 
\frac{1}{\rho(J^{\pi}_{i},E_{x})} 
\sum_{f} \sum_{XL} E_{\gamma}^{2L+1} \mathrm{f}_{XL} (E_{\gamma}) 
\label{eq:gamwid}
\end{equation}
%\end{widetext}

where $\rho(J^{\pi}_{i},E_{x})$ represents the density of initial levels and 
$\mathrm{f}_{XL} (E_{\gamma})$ is the photon strength 
for transition energy $E_{\gamma}$. The appropriate 
electric or magnetic character $X$ and multipolarity $L$ of the transition
is selected by spin and parity conservation. 
We have used the common practice of restricting the transition types to E1, M1 and E2
with no mixing, which leads to a single $XL$ choice for each final state.

For transitions into a bin of width $\Delta E$ in
the continuum part of the level scheme the density weighted average over final levels 
should be used:

%\begin{widetext}
%\begin{equation}
%\begin{split}
\begin{multline}
\langle \Gamma_{\gamma} (J^{\pi}_{i},E_{x}) \rangle = 
\frac{1}{\rho(J^{\pi}_{i},E_{x})} \sum_{f} \sum_{XL} \int_{E}^{E+\Delta E}
E_{\gamma}^{2L+1} \\
\times \mathrm{f}_{XL} (E_{\gamma}) \rho(J^{\pi}_{f},E_{x}-E_{\gamma}) \, dE_{\gamma}
\label{eq:gamwid2}
\end{multline}
%\end{split}
%\end{equation}
%\end{widetext}

Likewise the  average neutron width can be obtained by summation over all final states
of spin-parity $J^{\pi}_{f}$ and excitation energy $E_{x}-S_{n}-E_{n}$ in the final nucleus:

\begin{equation}
\langle \Gamma_{n} (J^{\pi}_{i},E_{x}) \rangle = 
\frac{1}{2 \pi \rho(J^{\pi}_{i},E_{x})} \sum_{f} \sum_{ls} T^{ls}(E_{n})
\end{equation}

where $T^{ls}(E_{n})$ is the neutron transmission coefficient, a function of neutron
energy $E_{n}$. The orbital angular momentum $l$ and channel spin $s$ 
are selected by spin and parity conservation for each final level.

% Create the reference section using BibTeX:
%\bibliography{basename of .bib file}

\end{document}